\documentclass[aps,pra,notitlepage,floatfix,amsmath,amssymb,superscriptaddress,twocolumn,showpacs]{revtex4-1}

\usepackage{graphicx}
\usepackage{epstopdf}
\usepackage{dcolumn}
\usepackage{bm}
\usepackage{amssymb}
\usepackage{setspace}
\usepackage{hyperref}
\hypersetup{%
  pdfpagemode=UseNone, %FullScreen,n
  pdfstartpage=1,
  pdfstartview=FitH,
  pdfmenubar=true,
  pdftoolbar=true,
  colorlinks = true,
  linkcolor=blue,
  citecolor=blue,
  bookmarksopen=false,
  urlcolor=black
}

\renewcommand{\equationautorefname}{Eq.}
\def\equationautorefname~#1\null{Eq. (#1)\null}

\newcommand{\ie}{i.\,e.,~}%
\newcommand{\eg}{e.\,g.\,~}%

\begin{document}

\title{Controlling the $s$-wave scattering length with non-resonant light:
  Predictions of an asymptotic model}

\author{Anne Crubellier}
\email{anne.crubellier@u-psud.fr}
\affiliation{Laboratoire Aim\'e Cotton, CNRS, Universit\'e Paris-Sud,
  Universit\'e Paris-Saclay, ENS Cachan, Facult\'e des Sciences
  B\^atiment 505, 91405  Orsay Cedex, France} 

\author{Rosario Gonz\'alez-F\'erez}
\email{rogonzal@ugr.es}
\affiliation{Instituto Carlos I de F\'{\i}sica
  Te\'orica y Computacional and Departamento de F\'{\i}sica At\'omica,
  Molecular y Nuclear, Universidad de Granada, 18071 Granada,
  Spain}

\author{Christiane P. Koch}
\email{christiane.koch@uni-kassel.de}
\affiliation{Theoretische Physik,  Universit\"at Kassel,
  Heinrich-Plett-Str. 40, 34132 Kassel, Germany}

\author{Eliane Luc-Koenig}
\email{eliane.luc@u-psud.fr}
\affiliation{Laboratoire Aim\'e Cotton, CNRS, Universit\'e Paris-Sud,
  Universit\'e Paris-Saclay,  ENS Cachan, Facult\'e des Sciences
  B\^atiment 505, 91405  Orsay Cedex,
  France}

\date{\today}
%:::::::::::::::::::::::::::::::::::::::::::::::::::::::::::::::::::::::::::
\begin{abstract}
A pair of atoms interacts with non-resonant light via its anisotropic polarizability. This effect can be used to tune the
scattering properties of the atoms. 
Although the light-atom interaction varies with interatomic separation
as $1/R^{3}$, the effective $s$-wave potential decreases more rapidly,
as $1/R^{4}$ such that the field-dressed scattering length can be
determined without any formal difficulty.
The scattering dynamics are essentially governed by the long-range
part of the interatomic interaction and can thus be accurately described by an asymptotic model [Crubellier et al., New J. Phys. 17, 045020 (2015)]. 
Here we use the asymptotic model to determine the field-dressed
scattering length from the $s$-wave radial component of a particular
threshold wave function. Applying our theory to the scattering of two
strontium isotopes, we calculate the variation of the scattering
length with the intensity of the non-resonant light. Moreover, 
we predict the intensities at which the scattering length becomes
infinite for any pair of atoms.
\end{abstract}
%::::::::::::::::::::::::::::::::::::::::::::::::::::::::::::::::::::::::::::::
% 34.50.Cx Elastic; ultracold collisions 
% 34.50.Rk Laser-modified scattering and reactions 
\maketitle
%-------------------------------------------------------------------------------
\section{Introduction}
\label{sec:intro}
%-------------------------------------------------------------------------------
Collisions of neutral atoms at very low temperatures are universally
described by a single parameter, the $s$-wave scattering length 
for bosons and unpolarized fermions or the $p$-wave scattering volume 
for polarized fermions. These parameters determine the strength of the 
contact potentials for all partial waves~\cite{Derevianko05,Idziaszek06}. 
Their values determine whether an ultracold gas is weakly or strongly interacting; and their sign renders the interaction to be effectively attractive or repulsive~\cite{Dalibard98}. 
This is important for example for the stability of Bose-Einstein
condensation or the stability of Fermi gases against collapse at high densities. Controlling the scattering length or the scattering volume has therefore long been a primary goal in quantum gas experiments~\cite{GiorginiRMP08,ChinRMP10,TheisPRL04,BlattPRL11,YamazakiPRA13,KillianPRL13}. 

While initial proposals suggested optical means for
control~\cite{FedichevPRL96,BohnPRA97,SlavaJCP01,CiuryloPRA05}, tuning the scattering length using a magnetic field and Feshbach resonances turned out to be more practical~\cite{GiorginiRMP08,ChinRMP10}. This requires, however, the existence of a hyperfine manifold on the atom and a sufficient width of the Feshbach resonance. In contrast, optical control of the scattering length is ubiquitous. It was demonstrated for narrow-line transitions that are found for example in alkaline earth atoms~\cite{BlattPRL11,YamazakiPRA13,KillianPRL13}. Despite the comparatively long lifetimes of the metastable states employed in these experiments, control was limited by non-negligible losses. 

Non-resonant light can also be used to tune the scattering
length~\cite{GonzalezPRA12,TomzaPRL14}. It couples to the
polarizability anisotropy of the collision complex and, for sufficiently
high intensity, modifies the scattering length~\cite{TomzaPRL14}. 
This variation is similar to the control of the scattering length by a
magnetic field near a Feshbach resonance~\cite{ChinRMP10}. In particular, the scattering length diverges when a bound level is located exactly at threshold. For non-resonant light control  this occurs when a shape 
resonance crosses the threshold to become bound or when 
the field-dressed $s$-wave potential is sufficiently 
deepened to accomodate an additional bound level. Remarkably, 
non-resonant light control is of universal character, independent of the frequency of the light and the energy level structure of the molecule, 
as long as the frequency remains far from any molecular resonance~\cite{GonzalezPRA12,TomzaPRL14}. 

Collisions at very low temperature are essentially governed by the long-range part of the inter-particle interaction. The scattering properties 
are therefore very well described by asymptotic models, which account
only for the asymptotic part of the interaction potential~\cite{GaoPRA98,GaoPRA01,GaoJPB03,CrubellierJPB06,GaoPRA09,LondonoPRA10,CrubellierNJP15a,CrubellierNJP15b}.
The asymptotic Hamiltonian describing non-resonant light control of a pair of atoms~\cite{CrubellierNJP15a,CrubellierNJP15b} 
is identical to that found for DC electric field control of the atom-atom interaction~\cite{MarinescuPRL98} as well as that for ultracold collisions of polar molecules~\cite{RoudnevJPB09,BohnNJP09}. All of these problems are governed by the anisotropic dipole-dipole interaction, which decreases as $1/R^{3}$ (where $R$ is the inter-particle separation)
and which introduces a coupling between partial waves of the 
same parity. 

For an isotropic potential which decreases asymptotically as $1/R^3$, it is well known that the scattering length cannot be defined~\cite{OMalleyJMP61}. 
In this case, the scattering phase shift at low energy cannot be expanded in powers of the wave number of the colliding particles since the threshold wave function contains a $\log{R}$ contribution in addition to the term proportional to $(R -\widetilde{a})$ which is used to define the scattering length
$\widetilde{a}$. However, the dipole-dipole interaction is anisotropic and only of 'quasi long-range' character~\cite{MarinescuPRL98} which does allow to  define the scattering length without any particular difficulty. 
'Quasi long-range' refers to the fact that the effective $s$-wave potential decreases as $1/R^4$ for large $R$, and only the potential for the higher partial waves contains a diagonal long-range contribution $\propto -1/R^3$~\cite{DebPRA01}. 

Here we investigate non-resonant light control of the scattering length. 
We show that the asymptotic model together with the nodal line technique~\cite{CrubellierEPJD99,PasquiouPRA10}, previously developed to study non-resonant light control of shape resonances~\cite{CrubellierNJP15a,CrubellierNJP15b}, 
can be extended to the control of the field-dressed $s$-wave scattering
length $\widetilde{a}$. 
To predict the dependence of the field-dressed scattering length on the non-resonant field intensity,
the asymptotic model only requires knowledge of the reduced mass, atomic polarizability and field-free $s$-wave scattering length.
The difficulty that we have to address here is the problem of degenerate coupled continua. 
We obtain $\widetilde{a}$ from a particular threshold wave function which has a linear variation in the $\ell=0$ partial wave channel and does not diverge in all other ones.

The paper is organized as follows:  First, in Sec.~\ref{sec:theory}, we review  the asymptotic model for an atom pair interacting with non-resonant light via the polarizability anisotropy~\cite{CrubellierNJP15a}. We 
present the Hamiltonian in two types of reduced units, best adapted to either the non-resonant field control or the dipole-dipole interaction.
In Sec.~\ref{sec:nodal-line}, we describe the method to calculate
scattering wave functions. In particular, in Sec.~\ref{subsec:scatt-length}, we discuss the asymptotic form 
imposed on the degenerate threshold wave functions. It is a judicious choice of this form that allows for determining the scattering length. For more details of the method, in particular an assessment of its validity and the optimal choice of the asymptotic boundary conditions, the reader is referred to Appendix~\ref{app:scatt}.
The discussion is based on the comparison between single-channel analytical calculations based on an extension of the Levy-Keller approach~\cite{LevyJMP63} and additional multi-channel numerical calculations.
In Sec.~\ref{sec:results}, we first apply our model to two strontium isotopes $^{88}$Sr and $^{86}$Sr 
and calculate the dependence of the intraspecies $^{88}$Sr-$^{88}$Sr and the interspecies $^{86}$Sr-$^{88}$Sr scattering lengths on the non-resonant light intensity. We then exploit the universality of ultracold collisions, captured by the asymptotic model, and predict the non-resonant light intensity at which the scattering length becomes infinite. We conclude in Sec.~\ref{sec:conclusion}. 
%
%-------------------------------------------------------------------------------
\section{Length and energy scales for anisotropic $R^{-3}$ interaction: Hamiltonian and reduced units}
\label{sec:theory}
%-------------------------------------------------------------------------------
%
%*****************************************************
%\subsection{Hamiltonian and reduced units}
\label{sec:hamil-ured}
%****************************************************
%
In the Born-Oppenheimer approximation, the Hamiltonian describing the nuclear relative motion of a pair of atoms interacting with a non-resonant laser field 
is given by  
\begin{equation}
  \label{eq:2D_Hamil}
  H =   T_R+\frac{\hbar^2{\mathbf{L}}^2}{2\mu  R^2}+V_g(R)
  -\frac{2 \pi I}{c}\left(\Delta\alpha(R)\cos^2\theta+\alpha_\perp(R)\right)\,,
\end{equation}
where $T_R$ and $\hbar^2 \mathbf{L}^2/2\mu R^2$ are the vibrational and rotational kinetic energies. $\mu$ denotes the reduced mass, $R$ the interatomic separation, $V_g(R)$ the interaction potential of the electronic 
ground state.  
The last term in the Hamiltonian~\eqref{eq:2D_Hamil} stands for the interaction 
with a non-resonant light field of intensity $I$, 
linearly polarized along the space-fixed $Z$ axis. $\theta$ 
denotes the polar angle between the interatomic axis and the laser 
polarization axis, and $c$ the velocity of light.
The perpendicular and parallel components of the 
polarizability tensor, $\alpha_\perp(R)$ and  $\alpha_\parallel(R)$,
are defined with respect to the interatomic axis, and have the dimension of a volume~\cite{CrubellierNJP15a}.
The polarizability anisotropy is given by 
$\Delta\alpha(R)=\alpha_\parallel(R)-\alpha_\perp(R)$.

In the asymptotic approximation, $V_g(R)$ is replaced by the van der Waals interaction  $-C_6/R^6$,  and the molecular polarizabilities are  expressed in terms of the polarizabilities of the two constituent atoms, $\alpha_1$ and $\alpha_2$, as follows:
$\alpha_\parallel(R)=\alpha_1+\alpha_2+4\alpha_1\alpha_2/R^3$ and 
$\alpha_\perp(R)=\alpha_1+\alpha_2-2\alpha_1\alpha_2/R^3$~\cite{Silberstein1,Silberstein2,Silberstein3}. 
The rovibrational Hamiltonian~\eqref{eq:2D_Hamil} then becomes
\begin{eqnarray}
  \label{eq:2D_Hamilbis}
  H &=&   T_R+\frac{\hbar^2\mathbf{L}^2}{2\mu  R^2}-\frac{C_6}{R^6}
    \\ \nonumber
  &-&	\frac{2 \pi I}{c}\left(( \alpha_1 + \alpha_2)+2 \alpha_1 \alpha_2 \frac{3\cos^2\theta -1}{R^3}\right)\,.
\end{eqnarray}
In the second line, the term $E_0=-2\pi I (\alpha_1 +\alpha_2)/c$ 
represents the lowering 
of the dissociation limit due to the interaction with the non-resonant field. The last term in the parenthesis  is very similar to the one involved in  dipole-dipole scattering in ultracold gases of atoms or molecules, with either permanent or field-induced  dipole moments. This is not surprising since the polarizability coupling describes nothing 
but the interaction of the two dipoles induced by the non-resonant field.
The correspondence between Eq.~\eqref{eq:2D_Hamilbis} and the standard dipole-dipole interaction for electric, respectively magnetic, dipoles becomes obvious by writing 
\begin{equation}
  \label{eq:corresp}
 \frac{4 \pi I}{c}\alpha_1\alpha_2\,\leftrightarrow\,\frac{1}{4\pi\epsilon_0}d_1d_2\,\leftrightarrow\,\frac{\mu_0}{4\pi}m_1m_2\,,
\end{equation}
where $d_{1,2}$ ($m_{1,2}$) denotes the magnitude of the electric (magnetic) dipole moments. 

In the Hamiltonian~\eqref{eq:2D_Hamilbis},
 the van der Waals potential $V_g(R)=-C_6/R^6$ represents the 'short-range' physics, 
which prevails when the atoms are rather close together, but still in the asymptotic region, whereas the long-range dynamics is governed by the dipole-dipole interaction induced by the non-resonant field. 
Thus, two length and energy scales can be defined in the problem. The first one,
well adapted to the interaction with a low-intensity non-resonant field, is
independent of the field intensity and characteristic of the short-range interaction. The other one is characteristic of the dipole-dipole interaction and allows to highlight the universal character of the dipolar scattering.

Analogously to  Ref.~\cite{CrubellierNJP15a}, we introduce reduced units (ru)~\cite{LondonoPRA10} to treat  the centrifugal term and the van der Waals interaction on an equal footing. The reduced units  are 
$R=\sigma x$,   $E - E_0 =\epsilon \,{\mathcal E}$, and  $I=\beta~{\mathcal I}$ for length, energy,  and laser intensity, 
respectively, with 
\begin{subequations}
\label{eq:scaling}
\begin{eqnarray}
\sigma & = & \left(\frac{2\mu C_6}{\hbar^2}\right)^{1/4}\,, 
\label{eq:sigma}\\ %\nonumber
\epsilon & = & \frac{\hbar^2}{2\mu\sigma^2}\,,
\label{eq:epsilon} \\ %\nonumber
\beta & = & \frac{c}{12\pi} \frac{\hbar^{3/2}C_6^{1/4}}{\alpha_1\alpha_2(2\mu)^{3/4}} 
=\frac{c\sigma^3\epsilon}{12\pi\alpha_1\alpha_2}\,.
\label{eq:beta}
\end{eqnarray}
\end{subequations}
These unit conversion factors  contain information specific to the atom pair, i.e., the reduced mass $\mu$, the van der Waals coefficient  $C_6$, 
and the atomic polarizabilities $\alpha_{1}$ and $\alpha_{2}$. In reduced units, 
the asymptotic Schr\"odinger equation reads
\begin{equation}
  \label{eq:asy}
  \left[-\frac{d^2}{dx^2} - \frac{1}{x^6} + \frac{\mathbf{L}^2}{x^2}
    - {\mathcal I}  \frac{\cos^2\theta -1/3}{x^3} - {\mathcal E}
  \right] f (x,\theta,\varphi) = 0\,,
\end{equation}
where $\varphi$ is the azimuthal angle and $f (x,\theta,\varphi) $ is
 the wave function in the asymptotic limit. 

Following Ref.~\cite{BohnNJP09} and using the equivalent dipoles defined by the relations \eqref{eq:corresp}, 
we define a second set of reduced units  $R=D\overline{x}$ and $E=E_D\overline{\mathcal E}$, characteristic of the dipole-dipole interaction and involving thus intensity-dependent conversion factors, 
\begin{subequations}
\label{eq:scalingdip}
\begin{eqnarray}
D & = & \frac{\mu}{\hbar^2}\frac{4\pi\alpha_1\alpha_2}{c}I\,, 
\label{eq:D}\\
E_D & = & \frac{\hbar^2}{\mu  D^2} = \frac{4\pi \alpha_1 \alpha_2 }{c}\frac{I}{D^3} \,.
\label{eq:ED} 
\end{eqnarray}
\end{subequations}
In these reduced units, the asymptotic Schr\"odinger equation reads
\begin{equation}
  \label{eq:asydip}
  \left[-\frac{d^2}{d\overline{x}^2} - \frac{\overline c_6}{\overline{x}^6} + \frac{L^2}{\overline{x}^2}
    - 6  \frac{\cos^2\theta -1/3}{\overline{x}^3} - 2{\overline{\cal{E}}}
  \right] f (\overline{x},\theta,\varphi) = 0\,,
\end{equation}
where $\overline c_6$ is the reduced strength of the van der Waals interaction 
\begin{equation}
  \label{eq:c6}
\overline c_6 =  2 \mu C_6/(\hbar^2 D^4) \,.
\end{equation}

The two sets of reduced units  are related as 
\begin{subequations}
\label{eq:corresp-units}
\begin{eqnarray}
D & = & \frac{\mathcal I}{6}\,\sigma, 
\label{eq:D-sigma}\\ 
E_D & = & \frac{72 }{\mathcal I^2}\,\epsilon . 
\label{eq:ED-epsilon}
\end{eqnarray}
\end{subequations}
Whereas in~\autoref{eq:asy}, the short-range van der Waals interaction is described by a universal term, it is the long-range dipole-dipole 
interaction which appears as universal in~\autoref{eq:asydip}.
The non-universal parameters of the two equations, $\overline c_6$ and ${\mathcal I}$, are related by
\begin{equation}
\label{eq:tunable}
\overline c_6=\frac{\sigma^4}{D^4}=\frac{6^4}{{\mathcal I}^4}\,.
\end{equation}

To give an example, the reduced units for $^{88}$Sr$_2$ are $\sigma=151~a_0$,
$\epsilon=86$~$\mu$K and $\beta=0.636$~GWcm$^{-2}$~\cite{CrubellierNJP15a,CrubellierNJP15b}. 
For non-resonant intensities $\mathcal I \le 40$~ru (25~GWcm$^{-2}$
for strontium), the ratio $D/\sigma\le 6.6$ 
results in an intensity-dependent unit of length $D\le 10^3~a_0$. For comparison, we will consider the heteronuclear dialkali molecules with the smallest  and  the largest  permanent electric dipole moment, which amount to  $0.56$~Debye (KRb) and  $5.5$~Debye (LiCs)~\cite{BohnNJP09,LepersPRA13}.
For the dipole-dipole interaction between two KRb or two LiCs molecules, the reduced unit of length is very large, $D_\mathrm{KRb}=4800$~a$_0$ or $D_\mathrm{LiCs}=6\times 10^5~a_0$, while the reduced unit of energy 
is very small $E_{D,\mathrm{KRb}}=120$~nK or $E_{D,\mathrm{LiCs}}=7$~pK. 
In order to observe, with strontium atoms in a non-resonant field,  effects similar as those encountered in the dipolar scattering of these polar molecules, an intensity-dependent unit of length $D$ of the same order of magnitude is needed. This implies huge non-resonant field intensities, i.e., 
$\mathcal{I}\approx 100$~ru (65~GWcm$^{-2}$) would be required to
obtain the same behavior as for $\mathrm{KRb}$, and
$\mathcal{I}\approx 3500$~ru  (2200~GWcm$^{-2}$) to mimic $\mathrm{LiCs}$. 
The short-range  van der Waals interaction between two dialkali
molecules in their %rovibronic 
absolute ground states is very small, with  $\overline c_6=7\times10^{-6}$~ru or $\overline c_6=9\times10^{-9}$~ru for KRb or LiCs, respectively. 
In these systems, the long-range dipole-dipole interaction prevails, 
such that the van der Waals interaction can be neglected in~\autoref{eq:asydip}, and 
the short-range physics taken into account by including a repulsive
wall~\cite{BohnNJP09}. This is in contrast to strontium where, for the
non-resonant light intensity that we consider, the van der Waals term
cannot be neglected.

%******************************************************************************
\section{Asymptotic model and nodal line technique}
\label{sec:nodal-line}
%******************************************************************************
%
%+++++++++++++++++++++++++++++++++++++++++++++++++++
\subsection{General description of the method}
\label{subsec:partial-wave}
%+++++++++++++++++++++++++++++++++++++++++++++++++++
%
To solve the asymptotic Schr\"odinger equation~\eqref{eq:asy}, we
first expand the wave function $f(x,\theta,\varphi)$
in terms of spherical harmonics $Y_{\ell}^m(\theta,\varphi)$
with fixed magnetic quantum number $m$ and the same parity, \ie even or odd values of $\ell$, 
due to the symmetry of the Hamiltonian~\eqref{eq:2D_Hamilbis}.
For practicality, the infinite sum over partial waves needs to be 
restricted to $\ell$ varying from $\ell_{min}=|m|$
to $\ell_{max}$ and includes $n=(\ell_{max}-\ell_{min}+2)/2$ channels.
Thus, a solution of~\autoref{eq:asy} is given by
\begin{equation}
  \label{eq:decomp}
  { f}(x,\theta,\varphi) = \sum_{\ell=\ell_{min}}^{\ell_{max}} y_{\ell}(x) Y_\ell^m(\theta,\varphi)\,,
\end{equation}
where the sum runs over either even or odd values of $\ell$, and  
$y_{\ell}(x)\equiv y_{\ell,m}(x)$ is the radial component of the solution
in the channel $\ell$ and $m$. For simplicity, the magnetic quantum number  $m$ is not specified in the 
radial part of wave function $y_{\ell}(x)$.
The amount of channels $n$ that needs to be included in~\autoref{eq:decomp} depends on the energy and on the laser intensity. 

This basis set expansion transforms the 
asymptotic Schr\"odinger equation~\eqref{eq:asy} into the following system of coupled equations 
\begin{equation}
  \label{eq:asyvect}
 \frac{d^2}{dx^2}{\bf y}(x) + ({\bf M}+ \mathcal E \openone)\cdot  {\bf y}(x)= 0\,,
\end{equation}
where ${\bf y}(x)$ is the vector  formed by the radial functions $y_{\ell}(x)$,
$\bf{M}$  is the matrix representation of  $\frac{1}{x^6} - \frac{\mathbf{L}^2}{x^2} +
\mathcal I \frac{\cos^2\theta-1/3}{x^3}$ in the basis of the spherical harmonics, and $\openone$ is the identity. 

In the present problem, the $n$ considered   channels correspond to
the same dissociation limit $E_0$ lowered by the non-resonant
field. For $\mathcal E <0$, the energy levels are quantized and
non-degenerate.  The continuous spectrum is $n$-times degenerate and
at each energy $\mathcal E \ge 0$, $n$ linearly independent solutions
are to be determined. We denote the physical wave functions, which are
a solution of  \autoref{eq:asyvect} and 
satisfy the boundary conditions, by a radial vector ${\bf z}(x)$, in
contrast to ${\bf y}(x)$ which is a general solution of \autoref{eq:asyvect}.
The wave functions are calculated here by inward integration of the
asymptotic Schr\"odinger equation, starting at a large separation 
$x_{max}$ with boundary conditions which depend on the energy $\mathcal E$. 
In the nodal line technique~\cite{CrubellierNJP15a} one replaces the interaction at very small interatomic separations by boundary conditions at the frontier between the inner and the asymptotic domains. 

In detail, the physical radial wave function, i.e., all partial wave components $z_{\ell}(x)$, must vanish at the so-called nodal line~\cite{VanhaeckeEPJD04,CrubellierEPJD99}, ${x}_{0}\equiv{x}_{0}({\mathcal E},\ell,\mathcal I)$, given by 
\begin{equation}\label{eq:nodalline}
{x}_{0}=x_{00}+A{\mathcal E}+B\ell(\ell+1)+C\mathcal I\,.
\end{equation}
The parameters $x_{00}$, $A$, $B$ and $C$ are characteristics for each 
atom pair. In particular, $x_{00}$ is the position of a node of the field-free
threshold $s$-wave  wave function; it is  determined
unambiguously by the field-free $s$-wave  scattering length in reduced units
\cite{CrubellierJPB06}. The parameter $A$ accounts (to first order) for the
energy dependence of the node position for  wave functions with 
$\ell=0$, whereas $B$ describes the shift of the node of the threshold wave functions induced by the centrifugal term for different  
partial $\ell$-waves. The last term in $x_{0}$    
accounts for the effects of the non-resonant field in the inner domain $x<x_{00}$ to first order in the field intensity.
For more details on these parameters and how to choose them, the reader is referred to Ref.~\cite{CrubellierNJP15a}.

The wave functions ${\bf z}(x)$ are determined in two steps.
First, a set of linearly independent particular solutions labeled ${\bf y}^j(x)$ satisfying  the asymptotic boundary conditions 
in the $n$ channels is obtained by inward integration.
The number of such solutions is equal to $n$ in the bound spectrum $\mathcal E <0$ and to $2n$ in the continuous spectrum $\mathcal E \ge 0$.
Second, the physical wave functions ${\bf z}(x)$ are constructed 
as linear combinations of the particular solutions ${\bf y}^j(x)$ such
that they fulfill the boundary conditions at short range, on the nodal
line. 

For $\mathcal E<0$, to obtain the energy of a bound state at a certain laser intensity $\mathcal I$, each of the $n$ particular solutions ${\bf y}^j(x)$ is related to a specific channel $\ell$ in which the asymptotic boundary
condition imposes an exponential decay, while in all other channels the radial functions vanish asymptotically. 
Imposing the boundary conditions at short range is equivalent to
making a function vanish that depends on energy and that is defined in
terms of the radial components $y^j_\ell(x_0)$ on the nodal lines. The
roots, i.e., the energies for which the function vanishes determine the bound levels,  providing the quantization of the bound spectrum, see the appendix in Ref.~\cite{CrubellierNJP15a}.
Analogously, to find at which intensity there is a bound state at a given energy, \eg  just below threshold, the zeros of a function of intensity have to be computed.

For $\mathcal E >0$, there exists an infinite number of sets of physical solutions ${\bf z}^j(x)$ that can be calculated.
For any specific problem, there is a most suitable choice for the $n$ linear combinations of the $2n$ particular solutions 
which is defined by their asymptotic behavior. If the initial conditions of the inward integration are chosen properly, 
the relevant physical property can be determined in a straightforward way.
To study the resonance structure of the continuum (Ref.~\cite{CrubellierNJP15a}), the asymptotic behavior in each channel is described by combinations of regular and irregular spherical Bessel functions~\cite{Friedrich98}.
For that problem we impose to the $2n$ particular solutions to be either regular or irregular spherical Bessel functions at $x_{max}$ in one channel and zero in all other ones. 
Among all possible sets of $n$ physical combinations,  
we choose the 'standard' form, in which the combination for a given
channel asymptotically contains a regular component in this channel
only and irregular components in all channels.
The $n$ physical solutions ${\bf z}^j(x)$ are such linear combinations
which vanish at the $\ell$-dependent node position $x_{0}$~\eqref{eq:nodalline}: these conditions allow for a direct determination of the reaction, scattering and  time delay matrices, ${\bf K}(\mathcal E)$, ${\bf S}({ \mathcal E})$ and ${\bf Q}(\mathcal E)$,
respectively,  as described in Ref.~\cite{CrubellierNJP15a}. The ${\bf Q}(\mathcal E)$ matrix is well-adapted
to analyze shape resonances, by studying the energy dependence of its lowest eigenvalue~\cite{CrubellierNJP15a}.

In ~\autoref{subsec:scatt-length} below, we show which particular solutions $\mathbf y(x)$ and which 
combinations $\mathbf z(x)$ are suitable for determining  the scattering length, 
whereas in appendix~\ref{app:scatt}, we study how the outer boundary 
conditions affect the convergence of the calculation of the field-dressed scattering length.

%
%+++++++++++++++++++++++++++++++++++++++++++++++++++++++++++
\subsection{Threshold wave functions and field-dressed scattering length}
\label{subsec:scatt-length}
%+++++++++++++++++++++++++++++++++++++++++++++++++++++++++++
%

In order to  determine the field-dressed scattering length  ${\widetilde a}(\mathcal I)$, 
we construct a wave function at threshold ($\mathcal E=0$)  which varies linearly with $x$ 
in the $\ell=0$ channel and vanishes asymptotically in all other channels.
We determine this wave function, as described above, by first constructing $2n$ particular solutions, denoted here as ${\bf f}_-^j(x)$ 
and ${\bf f}_+^j(x)$ with $1\le j \le n$. The asymptotic 
behavior of the non-zero component of ${\bf f}_+^j(x)$ is divergent, $f_{+,\ell_j}^j(x) \sim x^{\ell_j+1}$, 
whereas  $f_{-,\ell_j}^j(x)$ does not diverge, $f_{-,\ell_j}^j(x) \sim x^{-\ell_j}$. 
Several choices for the asymptotic form of $f_{\pm,\ell_j}^j(x)$ are possible. 
Here we will use pairs of analytical linearly independent threshold solutions of potentials $v^{\ell}_p(x)=\ell(\ell+1)/x^2-c_p/x^p$, 
chosen as approximations of the asymptotic \textit{diagonal} term of the 
matrix $-{\bf M}$ in the channel $\ell_j$ in~\autoref{eq:asyvect} 
(this choice is discussed in Appendix \ref{subsec:initial}, see in particular \autoref{tab:fct}).
The form of the physical wave functions ${\bf z}^j(x)$ 
is given by a linear combinations of a single asymptotically divergent solution and all $n$ non-divergent solutions:
\begin{equation}
  \label{eq:Afct}
  {\bf  z}^j(x) =  \sum_{j'=1}^n [\,\delta_{j',j}\, {\bf f}_+^{j'}(x)
  - \,{\bf \overline M}_{j'}^{j}(x_{0},x_{max})\,{\bf
    f}_-^{j'}(x)\,]\,. 
\end{equation}

The matrix ${\bf \overline M}$  (not be confused with ${\bf M}$) is
determined by the boundary conditions on the nodal lines and depends
both on nodal lines and on $x_{max}$. For a given atom pair, the nodal
lines are fixed; in the following, we omit the $x_{0}$
dependence. 
One has  ${\bf \overline M}({x_{max}})={\bf N}_{-}({x_{max}})\,.\,[{\bf
  N}_{+}({x_{max}})]^{-1}$, where, as in Ref.~\cite{CrubellierNJP15a},  
the matrices ${\bf N}_{+}(x_{max})$ and ${\bf N}_{-}(x_{max})$ are defined 
by their matrix elements $\left({\bf
    N}_{\pm}(x_{max})\right)_{\ell}^{j}={f}_{\pm,\ell}^{j}(x_0)$,
which are the values of all radial particular solutions on the nodal lines. 
The scattering length is obtained from the 
$s$-wave component of the physical solution  ${\bf z}^{j=1}(x)$, which in our notation corresponds to $\ell_j=0$ and which is of the form
\begin{equation}
\label{eq:longdif-fct}
z^{j=1}_{\ell=0}(x)  =  f^{j=1}_{+,\ell=0}(x) - \sum_{j'=1}^n {\bf \overline M}^{j=1}_{j'}(x_{max}) f^{j'}_{-,\ell=0}(x)\,.
\end{equation}
In \autoref{eq:longdif-fct}, the first term behaves asymptotically as $f^{j=1}_{+,\ell=0}(x) \rightarrow x$. The contributions to the sum of the radial components with $j' \ge 2$ vanish asymptotically at least as $1/x^2$, 
whereas for $j'=1$, $f^{j'=1}_{-,\ell=0}(x) \rightarrow 1$. 
Using the definition $\mathcal M(x_{max})\equiv {\bf \overline M}^{j=1}_{j'=1}(x_{max})$, one has, for sufficiently large $x_{max}$, 
\begin{equation}
z^{j=1}_{\ell=0}(x_{max})  \approx  x_{max} - {\mathcal M}(x_{max})\,, %\,\mathrm{when}\,\,x_{max} \rightarrow \infty\,,
\label{eq:longdif-xmax}
\end{equation}
with
\begin{equation}
\lim_{x_{max} \rightarrow \infty} {\mathcal M}(x_{max}) ={\widetilde a}({\mathcal I}) \,.%\,\,\mathrm{when}\,\,x_{max} \rightarrow \infty\,.
 \label{eq:longdif}
 \end{equation}
The field-dressed scattering length is unambiguously defined
by~\autoref{eq:longdif}, if  the limit $x_{max} \rightarrow \infty$
exists for  ${\mathcal M}(x_{max})$ and if it is independent of the
boundary conditions of  $f_{+,\ell}(x)$ and $f_{-,\ell}(x)$ at
$x_{max}$. 

The non-resonant field introduces a coupling between different partial
waves which vanishes asymptotically as $1/x^3$, so that the definition
of the scattering length becomes questionable. Indeed, for an
isotropic potential decreasing as $-1/x^p$, 
the limit  of $\tan \delta_\ell(k) /k^{2\ell+1}$ for $k\to 0$ does not exist for $2\ell+3 \ge p$, with $\delta_\ell(k)$ being the asymptotic elastic scattering  phase shift at the wavenumber $k$ of the partial wave $\ell$~\cite{Landau,OMalleyJMP61}.
A $\log k$ term appears in its expression, and a $\log x$ contribution in the threshold wave function. However, in the $s$-wave channel, there is no diagonal contribution of the non-resonant field induced interaction, and  
the diagonal potential of~\autoref{eq:asy} and~\autoref{eq:asydip} behaves asymptotically as $-1/x^6$. As emphasized in Refs.~\cite{MarinescuPRL98,DebPRA01}, 
the  asymptotic behavior of a dipole-dipole interaction, equivalent to the field induced term considered here, can be termed quasi long-range, since it corresponds to an effective potential $\propto -1/x^4$ for  $\ell=0$ 
and truly long-range effective potentials $\propto -1/x^3$ for all other partial waves.
The effective $s$-wave potential
becomes more attractive as the laser intensity $\mathcal I$ increases, and never exhibits a centrifugal barrier.
As a consequence, in spite of the $1/x^{3}$ dependence of the non-resonant field interaction, 
the definition of the field-dressed $s$-wave scattering length does not pose any formal difficulty. 

We analyze the validity of the outlined procedure to compute the field-dressed scattering in detail in appendix~\ref{app:scatt}.  
In particular, in~\autoref{subsec:Masym}, we provide an extension of the single-channel 
analytical two-potential approach originally developed by Levy and Keller~\cite{LevyJMP63,HinckelmannPRA71} 
to  determine the near-threshold elastic scattering phase shift in a long-range potential. 
For a potential $V(x)$ written as a sum of $1/x^p$ terms, we show that the expansion of $\mathcal M(x)$ 
in powers of $1/x$ contains a first term $\mathcal M_0$, accounting for the short-range interaction, 
and a $1/x$ term, depending only on $V(x)$ and on its separation in two terms in the two-potential approach. 
This expansion allows for straightforward determination of $\mathcal M_0$,
 \ie of the field-dressed scattering length. We have compared
the results of the single-channel Levy-Keller approach to systematic multi-channel numerical calculations, 
with a small number of channels, various values of $x_{00}$ and $x_{max}$ and different asymptotic 
boundary conditions for $f_{\pm,\ell}(x)$. The comparison between analytical and numerical 
results reveals that there exists an optimal pair of functions for the initialization of 
the inward integration in the $\ell=0$ channel, corresponding to a particularly rapid convergence of 
$\mathcal M(x_{max})$. This allows for the computation of the field-dressed scattering 
length using a properly chosen $x_{max}$, see~\autoref{subsec:fits}.

%
%-------------------------------------------------------------------------------
\section{Results}
\label{sec:results}
%-------------------------------------------------------------------------------
%

The interaction of an atom pair with non-resonant light modifies the effective  potential of the vibrational motion such that a scattering state may become bound~\cite{GonzalezPRA12,TomzaPRL14}. A bound state localized just at the dissociation limit for a particular non-resonant field intensity corresponds to a divergence of the scattering length as a function of the intensity. As a result, a non-resonant field can be used to  control of the  scattering length of a pair of colliding atoms. 

To illustrate this control, we consider the two isotopes of strontium,
$^{88}$Sr and $^{86}$Sr, with the largest natural abundance,  $68\%$ and $16\%$, respectively, and no nuclear spin.
The scaling factors~\autoref{eq:scaling} adapted to the van der Waals interaction for the $^{88}$Sr-$^{88}$Sr and $^{86}$Sr-$^{88}$Sr atom pairs are reported in Table I of Ref.~\cite{CrubellierNJP15a}.
We have chosen these atom pairs due to their very different field-free $s$-wave scattering lengths, which also corresponds to a different structure for shape resonances~\cite{LondonoPRA10}.
$^{88}$Sr has an intraspecies scattering length close to zero, ${\widetilde a}(\mathcal I=0)$=-2~$a_0$=-0.013~ru, and field-free shape resonances with $\ell=4,\,8,\,12,\dots$; whereas the interspecies scattering length of $^{86}$Sr-$^{88}$Sr is very large, ${\widetilde a}(\mathcal I=0)$=100~$a_0$=0.664~ru, and field-free shape resonances occur for $\ell=2,\,6,\,10,\dots$~\cite{GaoPRA09}. 
For $^{88}$Sr-$^{88}$Sr we derived 'realistic' intensity-dependent nodal lines from the wave functions calculated in a single channel 
diagonalization of the full Hamiltonian (Eq.~\eqref{eq:2D_Hamil}) introduced in Ref.~\cite{GonzalezPRA12}.
For $^{86}$Sr-$^{88}$Sr, for which there are no previous reliable theoretical or experimental data, we used 'universal' analytical 
parameters depending only on the $s$-wave scattering length~\cite{CrubellierNJP15a}.
 
In the following, we use the asymptotic boundary conditions BC24$^*$, defined in Appendix~\ref{subsec:initial}, as initial conditions for the inward integration of the asymptotic Schr\"odinger equation. 

%
%+++++++++++++++++++++++++++++++++++++++++++++++++++
\subsection{Scattering of  $^{88}$Sr atoms: Realistic nodal lines}
\label{subsec:zi}
%+++++++++++++++++++++++++++++++++++++++++++++++++++
%%%%%%%%%%%%%%%%%%%%%%%%%%%%% fig1 88Sr2 and table%%%%%%%%%%%%%%%%%%%%%%%%%%%%%%%%%%%%%%%

\begin{figure}[tb]
  \centering
  \includegraphics[width=0.99\linewidth]{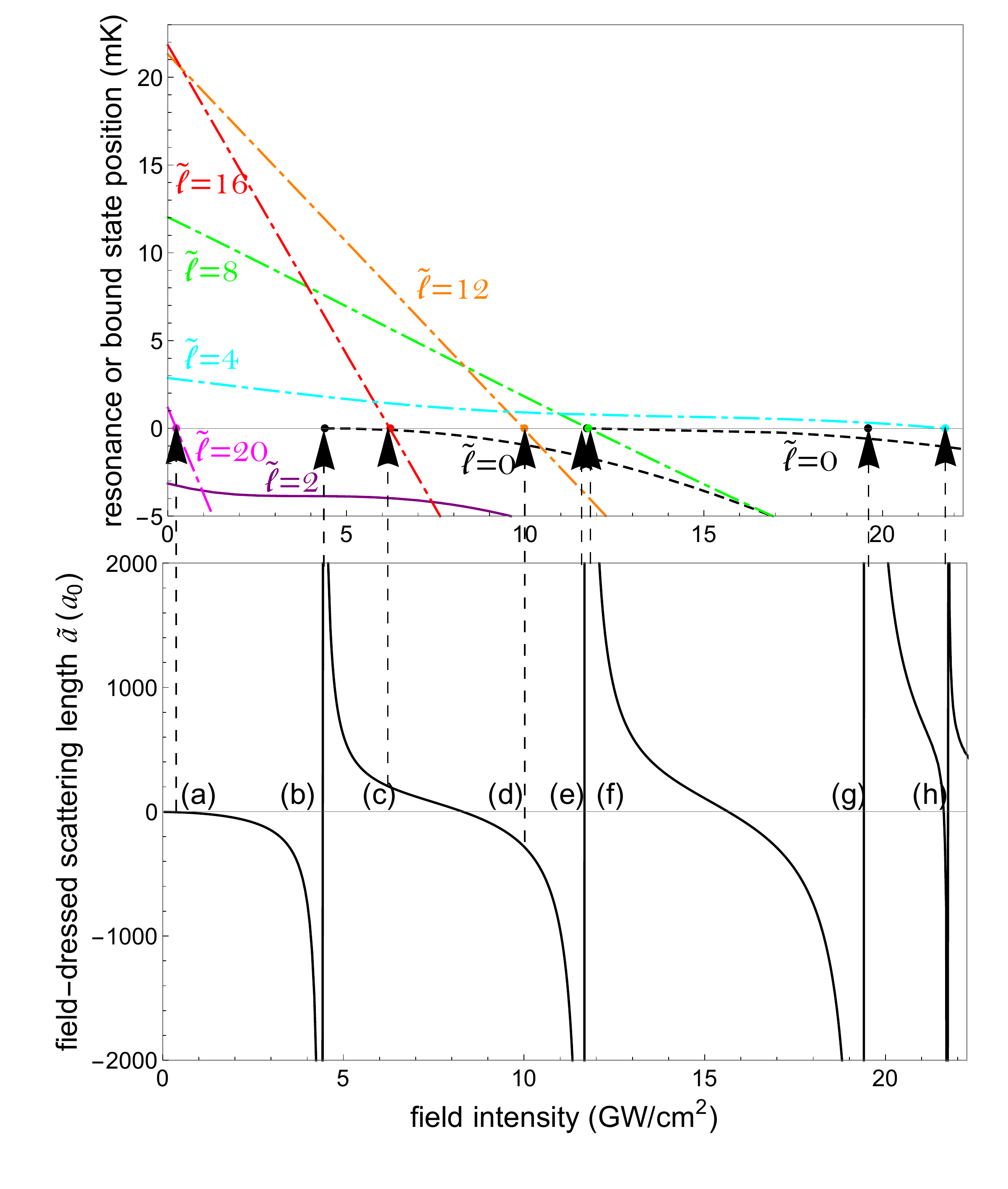}
  \includegraphics[width=0.99\linewidth]{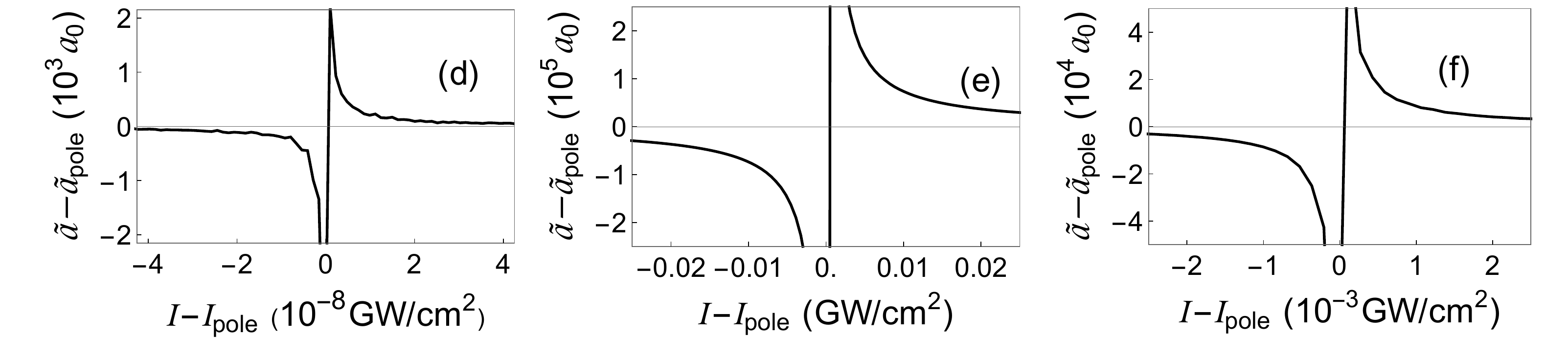}
  \caption{
    Comparison of the intensity-dependence of bound state energies and shape resonance positions (top) to the field-dressed scattering length of $^{88}$Sr atoms (middle and bottom). The resonances and bound states are characterized by diabatized labels ${\tilde \ell}$ (see text). Broad poles (b), (e) and (g) correspond to regular scattering states with ${\tilde \ell}=0$ becoming bound.  Narrow features (a), (c), (d), (f) and (h) are field-shifted shape resonances with ${\tilde \ell}= 20, 16, 12, 8$ and $4$, respectively. 
		Poles (d), (e) and (f), which do not clearly appear in the upper part, are shown in the 
		bottom diagrams, with subtraction of the background value 
		of scattering length $a_{1} + a_{2} \mathcal I_{pole}$ and of the resonance intensity $\mathcal I_{pole}$ 
		(see~\eqref{eq:pole}). The poles (a) and (c) are too narrow to obtain a precise shape.}
  \label{fig:scatt-zi}
\end{figure}
We have performed calculations with $11$ channels, using different $x_{max}$ in different channels. We have chosen $x_{max}$ such that we obtain, at this point and for the largest intensity considered, $\mathcal I=40$~ru (25~GWcm$^{-2}$), an $\ell$-purity close to one for each adiabatic eigenvalue.  
$x_{max}$ ranges from $x_{max}=20.9$~ru (3160~a$_0$) for  $\ell=0$ to $x_{max}=1.7$~ru (255~a$_0$) for $\ell=20$. 
The largest systematic error of the scattering length is obtained from the $\mathcal I^2/x_{max}$ coefficient of the expansion of $\mathcal M(x_{max})$ given in~\autoref{tab:fits} of the appendix. For  $\mathcal I=40$~ru and 
$x_{max}=20.9$~ru used for $\ell=0$, this error is of the order of $-0.04$~ru, which is absolutely negligible on the scale of the figures presented below. 

The field-dressed scattering length of $^{88}$Sr atoms is shown in \autoref{fig:scatt-zi} together with the position of shape resonances and bound states of $^{88}$Sr$_2$ as a function of the non-resonant light intensity 
(upper panel of~\autoref{fig:scatt-zi}.). 
The almost linear character of the energy dependence of the shape resonance positions is discussed in Ref.~\cite{CrubellierNJP15b}. All crossings between resonances or bound states are in fact anti-crossings. We introduce diabatized labels 
$\widetilde \ell$, where $\ell$ is the orbital momentum of the field-free states. 
The field-dressed scattering length exhibits divergences or poles at the non-resonant light intensities for which a bound state is located exactly at the dissociation threshold. This is indicated by the dashed arrows in~\autoref{fig:scatt-zi}. To determine the width $w$ of the poles, the field-dressed scattering length is fitted to the 
following expression 
\begin{equation}
\label{eq:pole}
f(\mathcal I)=a_{1}+a_{2}~\mathcal I+\frac{w}{(\mathcal I-\mathcal I_{pole})}\,,
\end{equation}
where we have introduced an intensity-dependent background  scattering length,  $a_{1}+a_{2}~\mathcal I$, and $\mathcal I_{pole}$ is the non-resonant light intensity at which the divergence occurs. The positions and widths of the singularities  of the scattering length are summarized in~\autoref{tab:zi}.
\begin{table}[tb]
  \begin{tabular}{|c|c|c|c|c|c|}
    \hline
      &$\widetilde{\ell}$& $\mathcal I_{pole}$ & $I_{pole}$          & $w$                   & $W$                   \\
 pole &label                     & ru &  GWcm$^{-2}$ & ru          & GWcm$^{-2}$           \\ \hline \hline
 (a)  &$20$ (S)                  & 0.368        & 0.234        & too small             & too small             \\ \hline
 (b)  &$0$                       & 6.93         & 4.41         & 3.43                   & 2.18                   \\ \hline
 (c)  &$16$ (S)                  & 9.77         & 6.21         & $\sim 10^{-12}$       & $\sim 10^{-12}$       \\ \hline
 (d)  &$12$ (S)                  & 15.7         & 9.96         & $ 2.21\,10^{-8}$ & $ 1.41\,10^{-8}$ \\ \hline
 (e)  &$0$                       & 18.4         & 11.7         & 7.62                   & 4.89                   \\ \hline
 (f)  &$8$ (S)                   & 18.5         & 11.8         & 0.0916                  & 0.0583                  \\ \hline
 (g)  &$0$                       & 30.5         & 19.4         & 13.6                  & 8.65                   \\ \hline
 (h)  &$4$ (S)                   & 34.2         & 21.7         & 0.641                  & 0.408                  \\ \hline
  \end{tabular}
  \caption{\label{tab:zi} 
  For the pair $^{88}$Sr~-$^{88}$Sr in  non-resonant light, 
features of the poles  of the field-dressed scattering length 
  (see~\autoref{fig:scatt-zi}): diabatized label $\widetilde \ell$, position and width in reduced units ($\mathcal I$ and $w$) and 
  in physical units ($I$ and $W$).  A pole denoted  by (S) is a shape resonance $\widetilde{\ell}>0$ becoming a 
  bound state as the intensity increases. A pole with 
  $\widetilde \ell=0$ is a supplementary  state, which appears  because  
 the  adiabatic $s$-wave potential becomes deeper.  The calculations are done for $n=$11 coupled channels with $\ell_{max}$=20. 
Realistic nodal lines (see text) are used.
} 
\end{table}

The new  bound states can be classified according to their rotational quantum number. Type (i) has $\widetilde \ell >0$ and corresponds to a shape resonance with  vanishing width that is pushed below threshold as the non-resonant field intensity increases.  The slope of the bound state energy as a function of the non-resonant field intensity is very large, and correspondingly the pole of the scattering length has very small width. Type (ii)  is a regular scattering state with $\widetilde \ell=0$ that becomes bound. This is due to the deepening of the field-dressed adiabatic $s$-wave potential, which can accommodate an additional bound state. For a supplementary bound state with $\widetilde \ell=0$, the width of the pole is very broad, because the energy of such a state remains  very close to the dissociation limit in a large range of intensities, see the dependence of the bound state energy  as a function of intensity in~\autoref{fig:scatt-zi} that starts tangentially to the threshold.

The pole structure of the field-dressed scattering length in~\autoref{fig:scatt-zi} appears to be very similar to that observed for magnetic Feshbach resonances~\cite{ChinRMP10}. However, there are two essential differences. First,  in a magnetic Feshbach resonance, there are at least one open and one closed channel with different dissociation limits, whereas in the present problem, all channels have the same dissociation limit. Magnetic Feshbach resonances arise from the coupling between a bound state in the closed 
channel and degenerate scattering states in the open one. In contrast, the divergence of the scattering length as a function of the non-resonant field appears when a shape resonance becomes bound. Second, for  magnetic Feshbach resonances, there is no equivalent to regular scattering states being accomodated as bound states in a modified $s$-wave potential.

The resonance structure in the energy dependence of the near-threshold cross section has been previously analyzed in Refs.~\cite{RoudnevJPB09,RoudnevPRA09}. In particular, for bosons, the resonances of the $\ell=0$, $m=0$ lowest adiabatic potential, which extend over a broad range of the cut-off radii modeling the short-range interaction, correspond to our type (i) poles. They produce an enhancement of the partial cross section averaged over 
the polarization direction by about two orders of magnitude.

%
%+++++++++++++++++++++++++++++++++++++++++++++++++++
\subsection{Interspecies scattering of $^{86}$Sr and $^{88}$Sr atoms:
 Universal nodal lines}
\label{subsec:86Sr88Sr}
%+++++++++++++++++++++++++++++++++++++++++++++++++++
% 
The power of the asymptotic model lies in the possibility to predict, at least roughly, the intensity-dependence of the field-dressed scattering length
for any atom pair. The field-free $s$-wave scattering length enters as
the only free parameter; it determines the universal nodal
lines~\cite{LondonoPRA10}. A universal nodal line refers
to~\autoref{eq:nodalline} with the three parameters $A$, $B$, and $C$
taking universal values. In particular, analytical formulas for the coefficients $A$ and $B$ in~\autoref{eq:nodalline}
are deduced from the universal model of Ref.~\cite{GaoPRA98}, which consists in a  $-1/x^{6}$ potential limited by an infinite repulsive wall at a distance
$x_{0_{G}} \rightarrow 0$. The shift of the node located at $x_{00}$, which is due to the contribution of the kinetic $A{\mathcal E}$ and centrifugal ${B\ell(\ell+1)}$ energies in the range $x_{0_{G}}\le x \le x_{00}$, can be evaluated using the WKB approximation~\cite{VanhaeckePhD}.  Specifically, 
the scattering length fixes the parameter $x_{00}$ in~\autoref{eq:nodalline}, and is used to estimate the universal coefficients  $A^G=-(x_{00})^7/8$ and
$B^G=(x_{00})^5/4$~\cite{LondonoPRA10}. For nodes at not too short range, the obtained values for $A^G$ and $B^G$  are comparable
%%% elk/ac: the nodes are never at long range
 those obtained from fitting to experimental data~\cite{CrubellierEPJD99,PasquiouPRA10}. There is also an analytical formula for the parameter $C^G$ accounting for the contribution of the non-resonant field at short-range%, see Eq. (13) in
~\cite{CrubellierNJP15a}.

%
%%%%%%%%%%%%%%%%%%%%%%%%%%% fig2 86Sr-88Sr %%%%%%%%%%%%%%%%%%%%%%%%%%%%%%%%%%%%%
\begin{figure}[tb]
  \centering
  \includegraphics[width=0.99\linewidth]{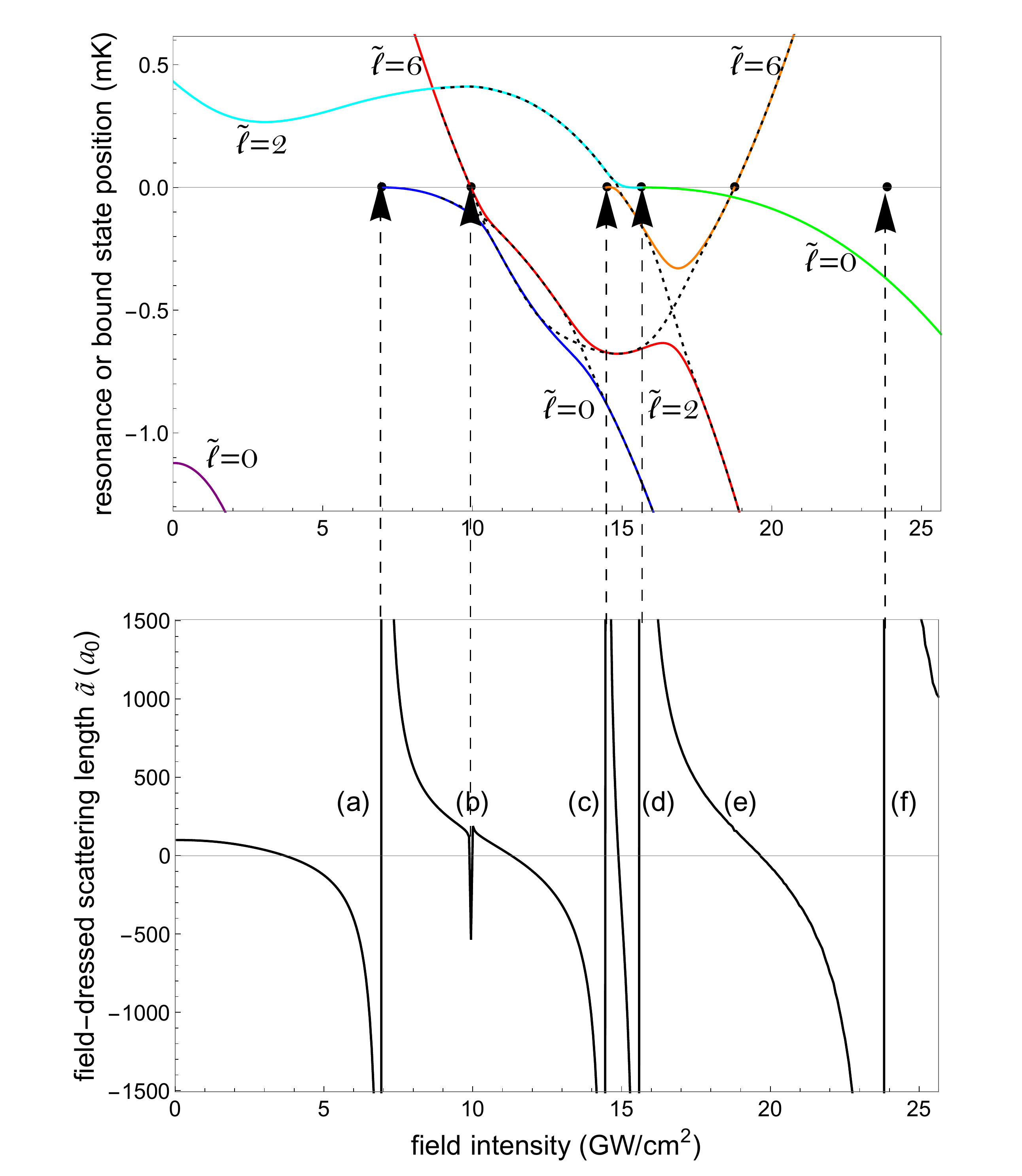}
  \includegraphics[width=0.99\linewidth]{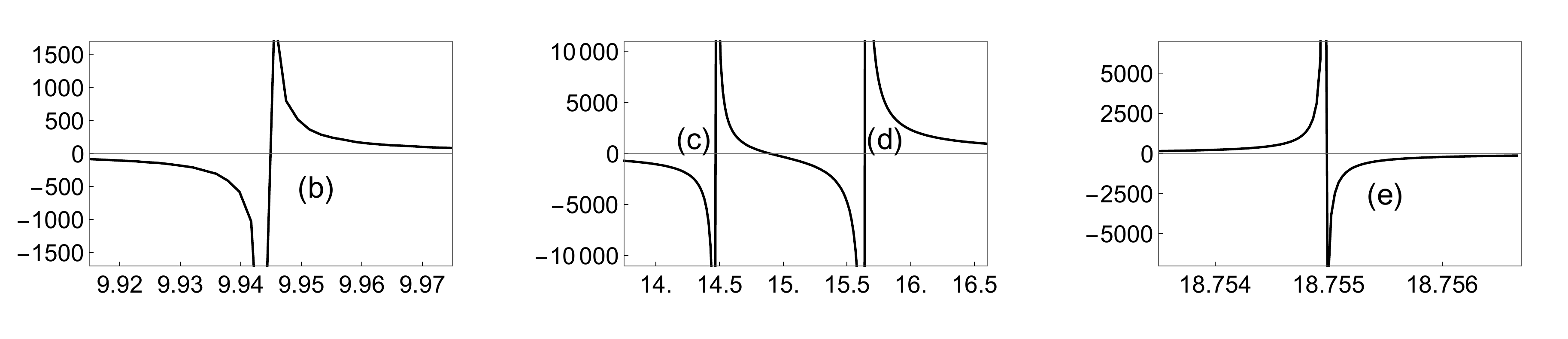}
  \caption{\label{fig:SLcontrol}
    Comparison of the intensity-dependence of bound
    state energies and shape resonance positions (top) to the
    field-dressed interspecies  
    scattering length of $^{86}$Sr and $^{88}$Sr atoms (middle and bottom). 
Due to the wide avoided crossings it is impossible to match colors and labels; diabatized curves are indicated by black dotted lines.
The broad poles of the scattering length (a), (d) and (f) correspond to regular scattering states with ${\tilde \ell}=0$ becoming bound, whereas narrow ones (b), (c) and (e) are shape resonances with ${\tilde \ell}= 6, 2$ and $6$, respectively, 
crossing the dissociation limit, as in~\autoref{fig:scatt-zi}.
%%% ac: it is not useful to repeat the axes labels
The narrowest poles are shown in the bottom diagrams with subtraction of the background scattering length for poles (b) and (e). In this part, axes labels are omitted, since they are exactly identical to those of the diagram just above. } 
\end{figure}
%%%%%%%%%%%%%%%%%%%%%%%%%%%%%%%%%%%%%%%%%%%%%%%%%%%%%%%%%%%%%%%%%%%%%%%%%%%%%%%%%%
%
%*********************************************************************************
\begin{table}[tb]
  \begin{tabular}{|c|c|c|c|c|c|}
    \hline
       &$\widetilde{\ell}$& $\mathcal I_{pole}$&     $I_{pole}$      & $w$          & $W$         \\
 pole  & label            & ru &  GWcm$^{-2}$   & ru   & GWcm$^{-2}$ \\ \hline \hline
 (a)   &$0$     &   10.9  &   6.99       &  5.29        &  3.39       \\ \hline
 (b)   &$6$ (S) &   15.5  &   9.94       &  0.0270      & 0.0173      \\ \hline
 (c)   &$2$ (S) &   22.6  &   14.5       &  3.74        & 2.40        \\ \hline
 (d)   &$0$     &   24.4  &   15.6       &  7.51        &  4.81       \\ \hline
 (e)   &$6$ (S) &   29.2  &   18.8       &  -0.00230    &  -0.00148   \\ \hline
 (f)   &$0$     &   37.2  &   23.8       &  18.9        &  12.1       \\ \hline
  \end{tabular}
  \caption{\label{tab:zi86-88} Same as~\autoref{tab:zi} but for the pair 
    $^{86}$Sr~-$^{88}$Sr. The calculations correspond to a $n=5$ coupled channel model (with
    $\ell_{max}$=8). Universal nodal lines (see text) are used.
        The exceptional shape of the resonance (e) in~\autoref{fig:SLcontrol} explains the sign of its width, see text.}
\end{table}
%
%*********************************************************************************
%
We now use the asymptotic model with these universal nodal lines to predict the intensity dependence of the field-dressed interspecies scattering 
length of $^{86}$Sr$-^{88}$Sr. The calculations were performed with
$n=5$ coupled channels, \ie  $\ell_{max}$=8. %,  and using universal
                                %nodal lines.  
For all channels, we have used  $x_{max}=40$~ru (6000~a$_0$). The results are presented in~\autoref{fig:SLcontrol}, and 
the positions and widths of the poles are summarized in~\autoref{tab:zi86-88}.

As for the intraspecies scattering of $^{88}$Sr atoms, there is a strong difference in the widths of the two types (i) and (ii) of poles
in~\autoref{fig:SLcontrol}. The avoided crossings among  bound states are very broad and widely avoided. Due to one of these anti-crossings, the ${\tilde \ell}= 6$ resonance crosses the threshold twice. As a consequence,
we obtain a negative width for the second  pole with ${\tilde \ell}= 6$ in the scattering length, see~\autoref{tab:zi86-88}. Note that  the scattering length is large and positive when there is a bound level close to threshold,
which in this case occurs for $\mathcal I<\mathcal I_{pole}$.

%+++++++++++++++++++++++++++++++++++++++++++++++++++
\subsection{Prediction of the field-dressed scattering length for any
  pair of atoms}
\label{subsec:div}
%+++++++++++++++++++++++++++++++++++++++++++++++++++
% 
%
%%%%%%%%%%%%%%%%%%%%%%%%%% fig3 any dimer %%%%%%%%%%%%%%%%%%%%%%%%%%%%%%%%%%%%%%%%%%
\begin{figure}[tb]
  \centering
  \includegraphics[width=0.99\linewidth]{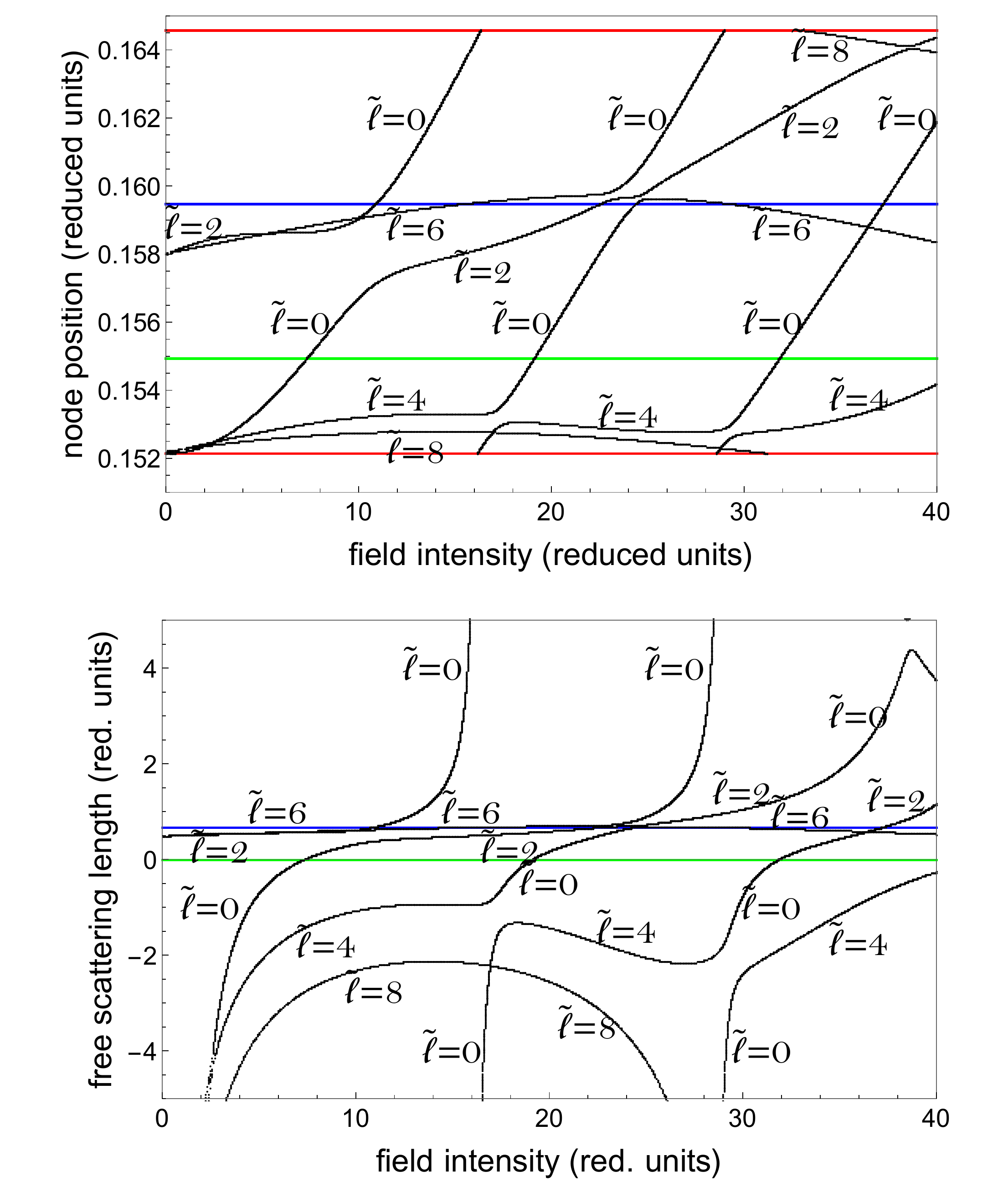}
  \caption{\label{fig:SL-poles} 
    Non-resonant light intensity, in reduced units, leading to 
    a divergence of the field-dressed scattering length. Each atom pair is characterized either by a node position $x_{00}$ of its field-free $s$-wave threshold wave function (upper panel) or by its field-free scattering length
    (lower panel), indicated by horizontal lines in either diagram. 
The red lines corresponds to an infinite field-free reduced scattering length, whereas the green and blue lines represent $^{88}$Sr-~$^{88}$Sr and $^{86}$Sr-~$^{88}$Sr, respectively.
A diabatic labeling is used. $\widetilde \ell>0$ describes a shape resonance becoming bound at the dissociation threshold and $\widetilde \ell=0$ denotes a regular scattering state that turns into a supplementary bound level in the  adiabatic field-dressed $s$-wave	potential. 
}
\end{figure}
%%%%%%%%%%%%%%%%%%%%%%%%%%%%%%%%%%%%%%%%%%%%%%%%%%%%%%%%%%%%%%%%%%%%%%%%%%%%%%%%%%%%%%%

%
%

The  asymptotic model with universal nodal lines can be used to predict the intensity-dependence of the scattering length for any pair of atoms, based on either the field-free scattering length or the node position of the field-free  $s$-wave threshold wave function. 
This is illustrated in~\autoref{fig:SL-poles} where, in the upper panel,  $x_{00}$ corresponds to the 6th node (counted from outside) of the $s$-wave threshold wave function.
The results have been calculated using $n=5$  channels, \ie $\ell_{max}=8$. 
We can read off from \autoref{fig:SL-poles} the non-resonant light intensities at which a new bound level appears at the dissociation limit, and, equivalently, a pole occurs in the field-dressed scattering  length $\widetilde a(\mathcal I)$: Any atom pair is characterized by a value of either  $x_{00}$ or $\widetilde a(\mathcal I=0)$. Drawing a horizontal line at this value, a pole in the field-dressed scattering length occurs when one of the black curves crosses the horizontal line. We have included the horizontal lines corresponding to infinite field-free scattering length as well as those for $^{88}$Sr-~$^{88}$Sr and $^{86}$Sr-~$^{88}$Sr in \autoref{fig:SL-poles} for illustration. 

For $x_{00}$, in the upper panel of~\autoref{fig:SL-poles}, we encounter 
two types of states at the dissociation threshold. For shape 
resonances ($\widetilde \ell > 0$) pushed below threshold (type (i) pole), atom pairs with similar values of $x_{00}$ have very different pole positions and the slope of $x_{00}$ as a function of the non-resonant field intensity is very small. This shows that the position, or energy, of a field-dressed shape resonance strongly depends on the short-range interaction. An exception is observed for the $\widetilde \ell =2$ states at $\mathcal I> 10$~ru and  the $\widetilde \ell =4$ states at $\mathcal I> 30$~ru, 
because the direct coupling between the $\ell=2$ and $\ell=0$
($\ell=2,\,4,$ and $6$) channels is very large. In contrast, for regular scattering states becoming bound in the modified $s$-wave potential (type (ii) poles), the field intensity characterizing the pole positions varies rapidly and almost linearly with  $x_{00}$. As a result, appearance of such a bound state depends mainly on 
the long-range properties of the field-dressed adiabatic $s$-wave potential and only to a lesser extent on the field-free scattering length. This can be seen in the lower graph of~\autoref{fig:SL-poles} where the $\widetilde \ell =0$ poles appear as almost vertical curves.

According to the field-free analytical model of Ref.~\cite{GaoEPJD04}, bound states with $\ell=4p$, $p \ge 0$ and integer, are located at threshold for an infinite scattering length, whereas for scattering lengths of $\widetilde a(\mathcal I=0)=0.48\,$ru and $x_{00}=0.158\,$ru bound levels with $\ell=2+4p$ are located at the dissociation limit. These predictions are approximately reproduced in our asymptotic model with universal nodal lines in the field-free case, cf.~\autoref{fig:SL-poles}.  

The node position of $x_{00}=0.1595$~ru is associated to interspecies
scattering of $^{86}$Sr and $^{88}$Sr atoms (blue vertical line in the
top panel of ~\autoref{fig:SL-poles}),
and~\autoref{fig:SL-poles} confirms the results
of~\autoref{tab:zi86-88} which were also calculated with universal nodal lines.
A field-free scattering length of $-0.013$~ru and $x_{00}=0.1549$~ru
describe intraspecies scattering of $^{88}$Sr atoms (green vertical line). The asymptotic model with  universal nodal lines predicts three regular $s$-wave scattering states becoming bound at  $\mathcal I \approx 7$, $16$  and $32$~ru, \ie very close to the non-resonant field intensities obtained  with realistic nodal 
lines for which the parameters were adjusted to reproduce the
field-free scattering length, see~\autoref{tab:zi}. This confirms the
slight dependence of the corresponding poles of the field-dressed
scattering length on the nodal lines. In contrast, for $\widetilde \ell=4$ and $\widetilde \ell=8$, the 
states obtained with universal nodal lines are never located 
at the dissociation limit, unlike the results obtained with realistic nodal lines in~\autoref{tab:zi}. As discussed above, the energies of field-free shape resonances strongly depend on the node position in the $\ell$-channel, 
\ie on the $B$ cofficient in~\autoref{eq:nodalline}, a sensitivity that increases for higher partial waves due to the factor $\ell(\ell +1)$. Our present findings are in line with Ref.~\cite{CrubellierNJP15a} where an overestimation of the field-free shape resonance positions with 
universal nodal lines was observed. 

The actual non-resonant field intensities that are required to control
the scattering length in strontium are rather large, 4.4$\,$GWcm$^{-2}$ and
7.0$\,$GWcm$^{-2}$ for intraspecies and interspecies scattering,
respectively. Control with a significantly smaller non-resonant field
intensity may be expected for atom pairs with a field-free scattering length between $0.48\,$ru and that of $^{86}$Sr and
$^{88}$Sr atoms, 0.664$\,$ru (or, correspondingly, $x_{00}$ between
0.158$\,$ru and 0.1595$\,$ru): 
Since $\widetilde a(\mathcal I=0)=0.48\,$ru is the value for which shape
resonances $\tilde\ell=2, 6, \ldots$ are located at threshold, a
comparatively small non-resonant field intensity is sufficient to
cross the  $\tilde\ell=2, 6$ curves in \autoref{fig:SL-poles} by the
horizontal line for an atom pair with a slightly larger field-free
scattering length. According to~\cite{LondonoPRA10}, field-free scattering lengths within this range are
found for the intraspecies scattering of $^7$Li in the ground singlet
electronic state and of $^{41}$K and $^{87}$Rb in either ground singlet
or lowest triplet state. Interspecies scattering of $^6$Li~-$^{40}$K,
$^6$Li~-$^{41}$K, $^7$Li~-$^{87}$Rb, $^7$Li~-$^{133}$Cs in the ground singlet
electronic state and   $^6$Li~-$^{7}$Li, $^{23}$Na~-$^{87}$Rb in the
lowest triplet state also exhibit field-free scattering lengths within
this range. The prospect of smaller non-resonant field intensities 
required than for the control of interspecies scattering in strontium
comes, however, with a grain of salt: The pole of the scattering
length will be narrow since a shape resonance, and not a regular
scattering state (with $\tilde\ell=0)$, is pushed below the
dissociation threshold.

In contrast, for atom pairs with negative field-free scattering length, the non-resonant field intensity must be at least larger than 2~ru.
The corresponding large intensities make the control of the
scattering length with a non-resonant field more challenging for
these atom pairs. For the example of intraspecies scattering of
$^{85}$Rb in the lowest triplet state, one would expect a $\tilde\ell
=0$ divergence at ${\mathcal I}$=3~ru (0.78~GWcm$^{-2}$) and a
$\tilde\ell =4$ divergence at ${\mathcal I}$=5~ru
(1.3~GWcm$^{-2}$). The $\tilde\ell=0$ pole will be broad and could be
exploited for non-resonant light control of the scattering. 

%
%---------------------------------------------------------------------------------------
\section{Conclusions}
\label{sec:conclusion}
%-------------------------------------------------------------------------------------
%

An asymptotic model has been used to analyze the scattering of two
ultracold atoms in a non-resonant laser field. It provides a realistic approximation to the field-dressed scattering length.
The pair of atoms interacts with the non-resonant light via its polarizability anisotropy such that the potential acquires  an  anisotropic $1/R^{3}$ asymptotic behavior. The asymptotic Hamiltonian is formally equivalent to the one describing the anisotropic dipole-dipole scattering in ultracold gases 
of atoms or molecules. While for a potential with an isotropic $1/R^{3}$ long-range dependence, the scattering length cannot be defined, the anisotropy induced by non-resonant light does not introduce a direct $1/R^{3}$ interaction term in the $s$-wave channel. 
In second order perturbation theory  the effective $s$-wave potential decreases  as $1/R^{4}$, 
and  the scattering length can be defined without any  formal difficulty. 
The field-dressed scattering length is determined from a threshold wave function 
which shows a linear variation with $R$ in the $s$-wave channel and does 
not diverge asymptotically for all higher partial waves.

In more detail, from the zero-energy scattering wave function, 
we have defined a quantity $\mathcal M(x)$ which tends to the scattering length for large $x$. 
We have first determined  $\mathcal M(x)$
analytically for a single channel model and a potential given by a multipole expansion. 
We predict in this way the expansion of $\mathcal M(x)$ in powers of $1/x$ 
for various choices of reference functions, with different asymptotic behavior.
In particular, a suitable choice of the reference functions allows to cancel the $1/x$ term. 
Then, for various atom pairs (\ie various $x_{00}$ or various field-free scattering length), 
we have calculated $\mathcal M(x)$ numerically
for a range of values of $x_{max}$. Fitting $\mathcal M(x_{max})$ has allowed us 
to deduce an expansion of $\mathcal M(x_{max})$ into a series of $1/x_{max}$. 
We have found the results of the numerical fits to agree well the analytical results 
when considering an expansion of both the $s$-wave adiabatic potential and 
the non-adiabatic coupling terms in powers of $1/x$, the latter being clearly non-negligible. 
The analysis has allowed us to estimate the uncertainty of the calculated 
scattering length when it is obtained from a single inward 
integration with judicious asymptotic boundary conditions at a value of $x_{max}$ which is not too large.

 We have shown that the scattering properties of an atom pair can be controlled by tuning the non-resonant field 
intensity. The scattering length diverges each time a field-dressed bound state reaches the threshold. We have encountered two types of divergences, with very different properties. On one hand, narrow divergences appear when a shape resonance with partial wave $\ell>0$ becomes bound. The corresponding non-resonant field intensity strongly depends on the interaction of the atom pair in the inner domain. On the other hand, broad divergences appear when a deepening of the  adiabatic field-dressed $s$-wave potential results in additional bound levels in the field-dressed $s$-wave channel. 
The corresponding intensities vary approximately linearly with the node position $x_{00}$ which replaces the inner part of the potential. 
The periodicity of the intensities at which the successive additional levels appear
is almost independent of the atom pair, since it is a characteristic of the 
long-range part of the field-induced interaction~\cite{RoudnevJPB09}. 

To illustrate the validity of the asymptotic method and the capability of non-resonant light to control the scattering length, we have considered the intraspecies scattering of $^{88}$Sr atoms and the interspecies scattering of $^{86}$Sr and $^{88}$Sr atoms as prototype systems. They have, respectively, a small and large  field-free scattering length and therefore  very different scattering properties. In contrast to shape resonances, where the intensity dependence of the resonance position was found to be rather different for the two cases~\cite{CrubellierNJP15a,CrubellierNJP15b}, the intensity dependence of the field-dressed scattering length turns out to be very similar, in particular for broad poles of the scattering length. This is explained by the fact that these poles occur when a regular scattering state becomes bound which is dictated almost exclusively by the field-induced dipole-dipole term, independently of the short-range interaction. 
In addition, we have considered an arbitrary pair of atoms, characterized only by the field-free scattering length, and we predict, in a completely general way, at which field intensity the field-dressed scattering length exhibits a divergence. This allows to estimate the non-resonant light intensity required to tune the scattering length for a given atom pair, provided the field-free scattering length is known. 

In future work, it will be important to consider the scattering of fermionic atoms. 
For a pair of  polarized fermionic atoms colliding in odd parity waves only, all channels involve an asymptotic $1/R^{3}$ potential and,   at ultracold temperature,  the usual $p$-wave scattering volume describing the collision cannot be defined.  
In a forthcoming paper~\cite{Crubellier16}, the theoretical tools  developed here for $s$-wave scattering will be adapted to describe the near threshold $p$-wave scattering of two fermionic atoms in a non-resonant laser field.

%--------------------------------------------------------------------------
\begin{acknowledgments}
Laboratoire Aim\'{e} Cotton is 
"Unit\'e mixte UMR 9188 du CNRS, de l'Universit\'e Paris-Sud,
de l'Universit\'e Paris-Saclay et de l'ENS Cachan", member of the
"F\'{e}d\'{e}ration Lumi\`{e}re Mati\`{e}re" (LUMAT, FR2764) and of
the "Institut Francilien de Recherche sur les Atomes Froids" (IFRAF).
R.G.F. gratefully acknowledges financial support by the Spanish project 
FIS2014-54497-P (MINECO), and by the Andalusian research
group FQM-207. 
\end{acknowledgments}
\appendix
%
%---------------------------------------------------------------------------------------
\section{Computation of the field-dressed scattering length}   \label{app:scatt} 
%---------------------------------------------------------------------------------------
%.

In this appendix, we discuss in detail how to compute the
field-dressed scattering length. The analytical functions chosen to
initialize the inward integration, by boundary conditions (BC) at $x_{max}$, are specified in~\autoref{subsec:initial}. 
In~\autoref{subsec:Masym} we suggest an extension of the single-channel approach by Levy and Keller~\cite{LevyJMP63,HinckelmannPRA71} to the threshold
wave function in order to calculate analytically the asymptotic phase-shift for near threshold elastic scattering in a long-range potential written in a multipole expansion. 
Using this extension, we obtain analytical asymptotic expansions of ${\mathcal M}(x_{max})$ in powers of $1/x_{max}$ for different asymptotic boundary conditions.
The adiabatic $s$-wave potential and the non-adiabatic contribution are examined in~\autoref{subsec:adiab}. 
For a small number of channels, we perform systematic calculations of ${\mathcal M}(x_{max})$   
and numerical fits to obtain the coefficients of the ${\mathcal M}(x_{max})$ 
expansion in powers of $1/x_{max}$, which are compared to the analytical results in \autoref{subsec:fits}. 

%
%*******************************************************************************
\begin{table*}[tb]
  \begin{tabular}{|l|c|c|c|c|c|c|}
    \hline
label &function&$p$&$\alpha$      &$\rho$               & $\gamma$  &  $q$            \\ \hline \hline
  BC2 &$x^{\ell+1}$&$0$&$-$           & $-$                 & $-$         &  $0$        \\
	    &$x^{-\ell}$&   &              &                     &           &              \\ \hline
 BC26 &$\gamma\sqrt{x}\,J_{\mp \alpha}(\rho)$& $6$ &$(2\ell +1)/4$ & $\sqrt{c_p}/(2x^2)$ & $(\sqrt{c_p}/4)^{\pm \alpha}\,\Gamma(1\mp \alpha)$&$-4$\\ \hline
 BC24 &$\gamma\sqrt{x} \,J_{\mp \alpha}(\rho)$ & $4$& $(2\ell +1)/2$&$\sqrt{c_p}/x$& $(\sqrt{c_p}/2)^{\pm \alpha}\,\Gamma(1\mp \alpha)$&$-2$      \\ %\hline 
\hline  %\hline
  \end{tabular}
  \caption{\label{tab:fct}  Analytical expression (column 2) of the pairs of linearly independent functions used either as asymptotic boundary conditions (BC, labeled in column 1) at $x_{max}$ in the channel $\ell$, 
or as reference pair of $\ell=0$ functions in the analytical model ($\varphi$ and $\psi$ being respectively the divergent and the non-divergent ones). The symbols $\alpha$, $\rho$ and $\gamma$ which appear in column 2 are given in columns  4, 5 and 6.
These functions are analytical solutions at threshold ($\mathcal E=0$) of the single $\ell$-channel Schr\"{o}dinger equation with the potential $v^{\ell}_p(x)=\ell(\ell+1)/x^2 - c_p/x^p$~\cite{MoritzPRA01}, where $p$ is given in the third column.  
The functions behave asymptotically as $x^{\ell+1}[1+O(c_p x^q)]$ and $x^{-\ell}[1+O(c_p x^q)]$ for the divergent (upper sign or upper line) and non-divergent  (lower sign or lower line) case, respectively, with  $q$ given in column 7. For free spherical waves, labeled by BC2 with $p=0$ and $c_p=0$, the functions are everywhere equal  to their asymptotic limit. 
}  
\end{table*}
%*******************************************************************************
%

%
%+++++++++++++++++++++++++++++++++++++++++++++++++++++++++++
\subsection{Asymptotic boundary condition for the inward integration}
\label{subsec:initial}
%+++++++++++++++++++++++++++++++++++++++++++++++++++++++++++
%

The pairs of  analytical functions defined in~\autoref{tab:fct} 
provide the asymptotic boundary conditions at $x_{max}$ in the channel $\ell$, \ie  initial conditions for inward integration in channel $\ell$. They are solutions of the Schr\"odinger equation associated to the asymptotic potential  $v^{\ell}_{p}(x)= \ell(\ell+1)/x^2 - c_p/x^p$, with $p > 2$, for zero energy~\cite{OMalleyJMP61,MoritzPRA01}. 
We present several choices~\footnote{In principle, any pair of functions with different logarithmic derivatives at $x_{max}$ could be used for each $\ell$ to obtain a set of solutions, but the asymptotic behavior of the solutions would in general be difficult to analyze.},   
generally expressed in terms of Bessel functions, which 
behave asymptotically as $x^{\ell+1}\,[1 + O(c_p x^{q})]$ and $x^{-\ell}\,[1 + O(c_p x^{q})]$, respectively. 

The labels are chosen as follows. The boundary conditions BC2 use $c_p=0$ and  $p=0$ such that  $v^{\ell}_{p}(x)$ reduces to the centrifugal term. The centrifugal term plus the van der Waals contribution, \ie $p=6$, is labeled by  BC26. The case BC24 with $p=4$ is well adapted to represent the effective field-dressed $s$-wave potential in a multi-channel description,  because both the adiabatic potential and the non-adiabatic couplings vanish asymptotically as $1/x^4$, see~\autoref{subsec:adiab}.

%
%+++++++++++++++++++++++++++++++++++++++++++++++++++++++++++
\subsection{ Two-potential approach to analytically determine the asymptotic behavior of ${\mathcal M}(x_{max})$ }
\label{subsec:Masym}
%+++++++++++++++++++++++++++++++++++++++++++++++++++++++++++
%
%*******************************************************************************
\begin{table*}[tb]
  \begin{tabular}{|l|c|c|c|}
    \hline
		label & $V_f(x)$ & $V_p(x)$& $\mathcal M(x)$ \\
		\hline
		\hline
		BC2   & $0$        & $-c_4/x^4-c_5/x^5-c_6/x^6$   & ${\mathcal M}_0+{c_4}/{x}+(c_5+c_4{\mathcal M}_0)/{x^2} +O({1}/{x^3})$ \\
		\hline
		BC26  &  $-1/x^6$  & $-c_4/x^4-c_5/x^5 -(c_6-1)/x^6$  & ${\mathcal M}_0+{c_4}/{x}+(c_5+c_4{\mathcal M}_0)/{x^2} +O({1}/{x^3})$ \\
		\hline
		BC24 &   $-c_{4f}/x^4$ & $ -(c_4-c_{4f})/x^4-c_5/x^5-c_6/x^6$ & ${\mathcal M}_0+(c_4-c_{4f})/{x}+(c_5+(c_4-c_{4f}){\mathcal M}_0)/{x^2} +O({1}/{x^3})$\\
		\hline
		\end{tabular}
		\caption{\label{tab:M(x)} Analytical expansion in powers of $1/x$ of $\mathcal M(x)$ (column 4) which characterizes the $s$-wave threshold solution $u(x)$~\autoref{eq:systLK} of the single-channel Schr\"{o}dinger equation in the potential $V(x)=V_f(x)+V_p(x)=-c_4/x^4-c_5/x^5-c_6/x^6$ when $u(x)$ is written in terms of two reference functions (labeled in column 1), which are threshold solutions of the potential $V_f(x)$ (column 2). $V_p(x)$ is given in column 3 and the expression of the reference functions is given in~\autoref{tab:fct}.}
		\end{table*}
%**********************************************************************************
%
We extend the two-potential method suggested originally by Levy and Keller \cite{LevyJMP63,OMalleyPR64,ShakeshaftJPB72} to determine, in a single-channel approach, the asymptotic behavior of ${\mathcal M}(x_{max})$ for the threshold $s$-wave solution of the Schr\"odinger equation involving an asymptotic  potential $V(x)=-c_4/x^4-c_5/x^5-c_6/x^6$. Following this method, a second potential $V_{f}(x)$ is defined to determine a pair of reference functions, $\varphi(x)$ and $\psi(x)$, which are linearly independent, analytical threshold solutions of this potential.
According to Ref. \cite{LevyJMP63}, the $s$-wave solution at threshold of the Schr\"odinger equation with the potential $V(x)$ can be written as
\begin{equation}
\label{eq:systLK}
u(x)=A(x)\left[\varphi(x)-\psi(x)\,{\mathcal M}(x)\right]
\end{equation}
with the condition 
\begin{equation}
\label{eq:systLK_2}
\frac{du(x)}{dx}=A(x)\left[\frac{d\varphi(x)}{dx}-\frac{d\psi(x)}{dx}\,{\mathcal M}(x)\right]\,.
\end{equation}
The Schr\"odinger equation for $u(x)$ with the condition~\eqref{eq:systLK_2} is equivalent to
\begin{subequations}
%\label{eq:solLK}
\begin{eqnarray}\label{eq:solLK-M}
\frac{d\,{\mathcal M}(x)}{dx}&=-\frac{V_p(x)}{W}\left[\varphi(x)-\psi(x)\,{\mathcal M}(x)\right]^2\, ,\qquad \\
\frac{1}{A(x)}\frac{d\,{A}(x)}{dx}&=-\frac{V_p(x)\,\psi(x)}{W}\left[\varphi(x)-\psi(x)\,{\mathcal M}(x)\right] \, ,\qquad 
\label{eq:solLK-A}
\end{eqnarray}
\end{subequations}
with $V_p(x)=V(x)-V_f(x)$ and  $W$ the Wronskian of $\varphi(x)$ and $\psi(x)$.

%
%*******************************************************************************

Here, we choose $V_f(x)=0 $  or $V_f(x)=-c_p/x^p$ with $p\ge 4$ and we
use for $\varphi(x)$ and $\psi(x)$ the pairs of analytical functions $\ell=0$ defined in~\autoref{subsec:initial}.
Since  for each $p$, the asymptotic limit of the reference functions 
are pure centrifugal wave functions, the scattering length $\widetilde{a}$ is the limit of $\mathcal M(x)$ for 
$x \rightarrow \infty$ (see~\autoref{eq:longdif}). 
For a chosen pair of reference functions ($\varphi(x)$, $\psi(x)$) associated with a given potential $V_f(x)$, the asymptotic 
form of $\mathcal M(x)$ is expanded in powers of $1/x$.  The coefficients of this expansion are obtained analytically by 
identifying the terms  with the same power of $1/x$ in~\autoref{eq:solLK-M}. It is obvious that this asymptotic analysis does not determine the constant coefficient $\mathcal M_0$, which does not appear in the left-hand side of the equation. This is not surprising since this coefficient depends on the short-range part of $V(x)$. Conversely, the other terms depend only on the potentials $V_p(x)$ and $V_f(x)$. 
The analytical expansion of ${\mathcal M}(x)$, corresponding to the 
labels of~\autoref{tab:fct}, is reported in~\autoref{tab:M(x)}. 
To the order $O({1}/{x^3})$, ${\mathcal M}(x)$ does not depend on the coefficient of the van der Waals interaction $-c_6/x^6$.
The coefficient of $1/x$ 
depends only on the $1/x^4$-terms in $V_p(x)$ and $V_f(x)$. The convergence of the ${\mathcal M}(x)$ expansion can be then controlled by a proper choice of $V_f(x)$. For example, the case BC24 allows to cancel the $1/x$-term in the expansion of ${\mathcal M}(x)$ by imposing $c_{4f}=c_4$, \ie by using reference functions related to the total $-c_4/x^4$ contribution to $V(x)$.

As mentioned above, this asymptotic analysis does not determine the constant coefficient of the expansion $\mathcal M_0$, \ie the scattering length $\widetilde a$.
However, if one introduces the analytical expansion  of ${\mathcal
  M}(x)$ in powers of $1/x$ into~\autoref{eq:solLK-A}, one can obtain
the analytical expansion of $A(x)$ in powers of $1/x$ (including the
$\mathcal M_0$ term) and then of the wave function $u(x)$ itself, which obviously does not depend on the choice of reference functions. One can finally compare the values of $\mathcal M_0$ corresponding to different reference pairs: It appears that the value of $\mathcal M_0$ does not depend on the choice of the reference pair.

%
%+++++++++++++++++++++++++++++++++++++++++++++++++++++++++++
\subsection{Adiabatic $s$-wave potential and non-adiabatic effects}
\label{subsec:adiab}
%+++++++++++++++++++++++++++++++++++++++++++++++++++++++++++
%
In~\autoref{tab:M(x)}, one can see that, when using the reference pair BC24 and $c_{4f}=c_4$, ${\mathcal M}(x)$ converges as $1/x^2$ instead of $1/x$ in the other cases. One may expect an increased convergence of the numerical multi-channel calculations when using the BC24 condition in the $s$-channel. To determine the optimal $c_{4}$ value,
we calculate the expansion in powers of $1/x$ of the effective $s$-wave potential in the matrix $-{\bf M}(x)$ in~\autoref{eq:asyvect}, 
and of the non-adiabatic contributions to this potential in a $n$-channel description.
 
For two coupled channels, with $\ell=0$ and $2$, an analytical expression 
can be found for the field-dressed adiabatic $s$-wave potential 
$v_{ad}^{\ell=0}(x)$ of the Hamiltonian~\autoref{eq:asy}. It is obtained as the lowest eigenvalue of the $2\times 2$ matrix $-{\bf M}(x)$ 
appearing in~\autoref{eq:asyvect} and given by the van der Waals
interaction plus a series of the form $\mathcal I^{q-2}/x^q$, 
\begin{equation}
v_{ad}^{\ell=0}(x)=-\frac{1}{x^6}-\frac{2}{135}\frac{\mathcal I^2}{x^4}-\frac{4}{8505}\frac{\mathcal I^3}{x^5}+\frac{58}{2679075}\frac{\mathcal I^4}{x^6}+O\left(\frac{\mathcal I^5}{x^7}\right)\,.\\
\label{eq:anal-ad}
\end{equation}
The diagonal contribution of the non-adiabatic coupling, due to the $x$-dependence of the corresponding adiabatic eigenvector $\Phi(x)$, \ie the 'kinetic energy' mean value  $\langle\Phi|\,d^2\,\Phi\,/dx^2\rangle$, reads
\begin{equation}
\label{eq:anal-n-ad}
v_{n-ad}(x)=-\frac{1}{405}\frac{\mathcal I^2}{x^4}-\frac{8}{25515}\frac{\mathcal I^3}{x^5}+\frac{64}{2679075}\frac{\mathcal I^4}{x^6}+O\left(\frac{\mathcal I^5}{x^7}\right)\,.
\end{equation}
Note the importance of this non-adiabatic term, which amounts to roughly 1/6 of the direct coupling term~\eqref{eq:anal-ad}.

For $n$=2, 3 and 4 channels, and for $\mathcal I=6,\,10,$ and $20$~ru, we have calculated the differences 
between the  eigenvalues of the matrix $-{\bf M}(x)$ and the analytical results for $n=2$. 
These differences have been least square fitted to a series of the form $\mathcal I^{q-2}/x^q$ and $q \ge 4$. The coefficients are always very small. The difference between the 
$n=3$ and $n=2$ cases  is even smaller, and between the $n=4$ and $n=3$ results almost negligible. This rapid stabilization of the adiabatic representation as the number of channels increases reflects the tridiagonal structure of the Hamiltonian matrix, with off-diagonal terms $\Delta\ell=\pm 2$. 

%
%+++++++++++++++++++++++++++++++++++++++++++++++++++++++++++
\subsection{Asymptotic behavior of ${\mathcal M}(x_{max})$: Numerical calculations}
\label{subsec:fits}
%+++++++++++++++++++++++++++++++++++++++++++++++++++++++++++

Here, we analyze the influence of the asymptotic boundary conditions
(BC) for the inward integration at $x_{max}$ given in~\autoref{subsec:initial} and of the number of coupled channels onto the field-dressed scattering length.
For $n=2$, 3, and  $4$ channels, \ie $\ell=0, 2$, $\ell=0, 2,4$ and
$\ell=0, 2, 4, 6$, respectively, we have performed a large number of numerical calculations in the interval $0.142152$~ru $\le x_{00}\le 0.152135$~ru, which corresponds to the field-free scattering length varying from  $-\infty$ to $+\infty$. Remember that each value of $x_{00}$ is associated to a specific pair of atoms.

\begin{table*}[tb]
  \begin{tabular}{|c|c|c|c|c|c||c|}
    \hline
${\mathcal I}$ &  label  & $V_f(x)$   & constant        & coefficient    & coefficient      & coefficient \\ 
 (ru)          &         &            & coefficient     & of $1/x_{max}$ & of $1/x_{max}^2$ & of $\mathcal I^2/x_{max}$\\ \hline \hline
               & BC2     & $0$        & $\widetilde {a}(\mathcal I,x_{00})$ (0.105964) & 0.64002 &  $\eta(\mathcal I,x_{00})$ & 0.017778\\ \cline{2-7}
     $6$       & BC26    &$-1/x^6$    & $\widetilde {a}(\mathcal I,x_{00})$   (0.105964)   & 0.64002    &  $\eta'(\mathcal I,x_{00})$  & 0.017778\\ \cline{2-7}
				       & BC24    &$-c_4/x^4$  & $\widetilde {a}(\mathcal I,x_{00})$   (0.105964)  & 0.1067     &  $\eta''(\mathcal I,x_{00})$  & 0.002964\\ \cline{2-7}
               & BC24$^*$&$-c_4^*/x^4$& $\widetilde {a}(\mathcal I,x_{00})$   (0.105964)  & 0.01781    &  $\eta''^*(\mathcal I,x_{00})$& 0.0004947 \\ \hline \hline
               & BC2     & $0$      & $\widetilde {a}(\mathcal I,x_{00})$   (-3.28493)  & 1.7779     &  $\eta(\mathcal I,x_{00})$   & 0.017779\\ \cline{2-7}
    $10$       & BC26    &$-1/x^6$  & $\widetilde {a}(\mathcal I,x_{00})$   (-3.28492)  & 1.778      &  $\eta'(\mathcal I,x_{00})$  & 0.01778 \\ \cline{2-7}
				       & BC24    &$-c_4/x^4$ & $\widetilde {a}(\mathcal I,x_{00})$   (-3.28492)  & 0.2962     &  $\eta''(\mathcal I,x_{00})$  & 0.002962 \\ \cline{2-7}
               & BC24$^*$&$-c_4^*/x^4$& $\widetilde {a}(\mathcal I,x_{00})$   (-3.28493)  & 0.0495     &  $\eta''^*(\mathcal I,x_{00})$& 0.000495\\ \hline \hline
               & BC2     &$0$       & $\widetilde {a}(\mathcal I,x_{00})$   (-1.77085)  & 7.1125     &  $\eta(\mathcal I,x_{00})$   & 0.017781\\ \cline{2-7}
    $20$       & BC26    &$-1/x^6$  & $\widetilde {a}(\mathcal I,x_{00})$   (-1.77086)  & 7.112      &  $\eta'(\mathcal I,x_{00})$  & 0.01778\\ \cline{2-7}
		           & BC24     &$-c_4/x^4$  & $\widetilde {a}(\mathcal I,x_{00})$   (-1.77086)  & 1.186      &  $\eta''(\mathcal I,x_{00})$  & 0.002965 \\ \cline{2-7}
				       & BC24$^*$&$-c_4^*/x^4$& $\widetilde {a}(\mathcal I,x_{00})$   (-1.77085)  & 0.198      &  $\eta''^*(\mathcal I,x_{00})$& 0.000495 \\ \hline
  \end{tabular}
  \caption{\label{tab:fits} The first three coefficients of the expansion of 
	$\mathcal M(x_{max})$ in powers of 1/$x_{max}$
	obtained by a fit performed with $x_{max}$ spanning the interval $[20$~ru$,500$~ru$]$, with different asymptotic boundary conditions (column 2, labeled BC and related to the potential $V_f$ given in column 3, see~\autoref{subsec:initial}) and three different non-resonant field intensities $\mathcal I$ (column 1), using three coupled channels ($\ell$=0, 2, 4). The calculations were performed using 150 or 200 values of $x_{00}$ (Eq.~\eqref{eq:nodalline}) chosen such that the field-free $s$-wave scattering length varies from $-\infty$ to $+\infty$. The table reflects the dependence on $x_{00}$ of these coefficients: they are either constant (coefficient of $1/x_{max}$, column 5) or else they present a characteristic shape (see~\autoref{fig:scatt-dim}) with three divergences (constant coefficient, column 4, and coefficient of $1/x_{max}^2$, column 6).
The constant coefficient $\mathcal M_0({\mathcal I},x_{00})$ varies with $x_{00}$ and $\mathcal I$, but not with the BC and 
is equal to the field-dressed scattering length $\widetilde {a}(\mathcal I, x_{00})$ (Eq.\eqref{eq:longdif}). 
Its particular value for $x_{00}=0.148741$~ru (\ie ${\widetilde a}({\mathcal I}=0)=0.786619$~ru)
is given inside brackets in column 4, to show the precision of its determination by the fit. 
For each value of $\mathcal I$ and for each type of initial conditions, the variation with $x_{00}$ of the coefficient of $1/x_{max}^2$, labeled $\eta(\mathcal I, x_{00})$ in column 6, depends on the boundary conditions (which is expressed by the different superscripts) but exhibits always a shape very similar to the one of $\widetilde {a}(\mathcal I, x_{00})$ : more precisely, it is related to $\mathcal M_0$ by a linear transformation (Eq.\eqref{eq:eta}). Column 7 shows the dependence on ${\mathcal I}$ of the coefficient of $1/x_{max}$.
} 
\end{table*}
%*****************************************************************************************************
%
In the multichannel asymptotic boundary conditions labeled BC2
(resp. BC26) in~\autoref{tab:fits}, we have used the pairs of functions BC2 (resp. BC26) of~\autoref{tab:fct} for any $\ell$. For the BC24 conditions, the BC24 functions were used in the $\ell=0$ channel and either the BC2 or BC26 pairs in the other channels, both leading to the same results.
For $v^{\ell=0}_{p=4}(x)$, we have considered either the adiabatic $s$-wave potential of the two-channel model~\eqref{eq:anal-ad}  with $c_4=2\mathcal I^2/135$ or this term plus the non-adiabatic contribution~\eqref{eq:anal-n-ad} with $c_4^*=7\mathcal I^2/405$. These two cases are referred to as BC24 and BC24$^*$, respectively.

\begin{figure}[tb]
  \centering
  \includegraphics[width=0.99\linewidth]{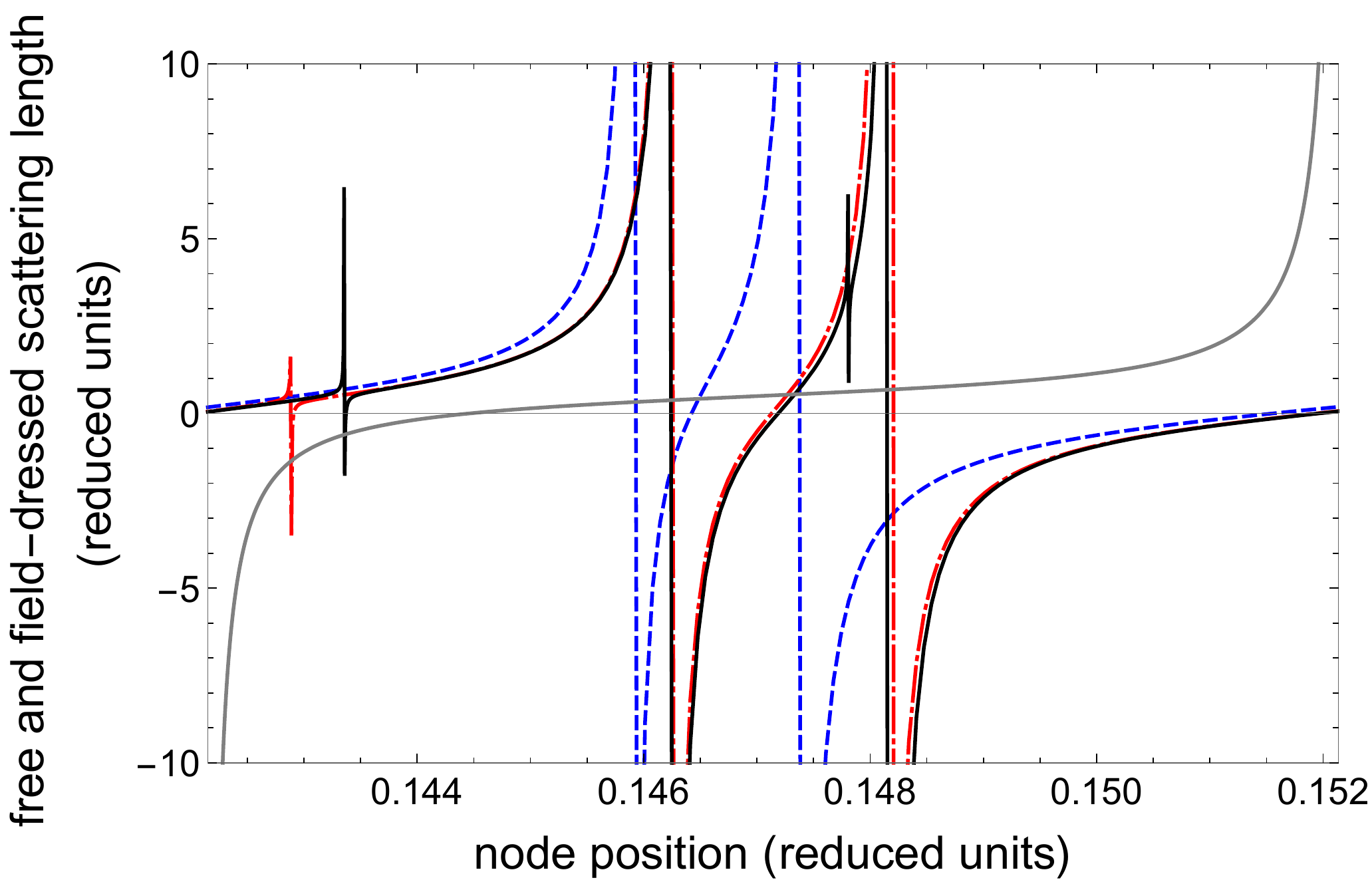}
	\caption{Field-dressed  $s$-wave scattering length for 
$n=2$ (blue dashed line), 
$n=3$  (red dot dashed line), and
$n=4$  (black solid line) channels,  and $\mathcal I=10$~ru.  
The field-free  scattering length (gray line) is also plotted.}
  \label{fig:scatt-dim}
\end{figure}

For each  $x_{00}$, we have calculated $\mathcal M(x_{max})$ for $20$~ru$ \le x_{max}\le 500$~ru, using the method described in~\autoref{subsec:scatt-length}. 
A fitting of $\mathcal M(x_{max})$ to a polynomial in $1/x_{max}$
provides the first coefficients of the expansion of $\mathcal
M(x_{max})$ in powers of $1/x_{max}$. In this way, we obtain the
dependence of these coefficients on $x_{00}$.
The results for three coupled channels, \ie $\ell=0, 2$ and $4$, are presented in~\autoref{tab:fits} for 
$\mathcal I =6, 10$, and $20$~ru. The coefficients are either $x_{00}$-independent (coefficient of $1/x_{max}$, column 5), or else they present a characteristic shape  (see~\autoref{fig:scatt-dim}) with several (three, here, since we consider three channels) divergences (constant coefficient, column 4, and 
coefficient of $1/x_{max}^2$, 
column 6), the position in $x_{00}$ of the divergences only depending on ${\mathcal I}$ and on the number of channels.

The constant term, ${\mathcal M}_0$, which is the field-dressed scattering length ${\widetilde a}({\mathcal I}, x_{00})$, 
only depends on $\mathcal I$ and on $x_{00}$, \ie on the field intensity and on the atom pair, but not on the boundary  conditions.  The accuracy of  this procedure is illustrated by values of the field-dressed scattering length for the specific choice of $x_{00}=0.148741$~ru (\ie ${\widetilde a}({\mathcal I}=0)=0.786619$~ru), see values in parenthesis of forth column of~\autoref{tab:fits}: A dependence on the boundary condition appears at most in the sixth digit.

The coefficient of $1/x_{max}$ is independent of $x_{00}$, but depends on the boundary conditions and is proportional to 
$\mathcal I^2$, see columns 5 and 7 of~\autoref{tab:fits}. The coefficients of $\mathcal I^2/x_{max}$ are identical for BC2 and 
BC26 ($0.01778$), smaller for BC24 ($0.00296$) and even smaller for BC24$^*$ ($0.00049$).  These differences
are due to the way the term $1/x^4$ is considered in the boundary conditions of the $s$-wave. 

To compare with the analytical results of \autoref{tab:M(x)}, we
examine the coefficients of the $\mathcal I^2/x_{max}$ factor in the $\mathcal M(x_{max})$ expansion (column 7 of~\autoref{tab:fits}). We first recall that an approximation for the $s$-wave field-dressed potential is $-c_{4}^*/x^4$, a sum of the adiabatic and non-adiabatic 2-channel contributions~\eqref{eq:anal-n-ad}. With the boundary conditions BC2 and BC26, which do not include a $1/x^4$ term, the coefficient of ${\mathcal I}^2/x_{max}$ is close to $c_{4}^*=0.01728$, as expected. 
When the direct coupling term $\propto 1/x^4$ between the $\ell=0$ and $\ell=2$ channels is included in the reference functions, the corresponding contribution disappears in the $\mathcal I^2/x_{max}$ coefficient of the $\mathcal M(x_{max})$ expansion: This explains why the difference between the boundary conditions BC2 and BC24 is $0.01482$ or $\approx 2/135$.
Analogously, a comparison of the BC24 and BC24$^*$ results shows that when the non-adiabatic couplings are  
accounted for, the $\mathcal I^2/x_{max}$ coefficient in the $\mathcal M(x_{max})$ 
expansion decreases by  $0.00247 \approx 1/405$.
These multi-channel results testing the influence  of the %introduction of 
asymptotic boundary conditions on the $\mathcal I^2/x_{max}$ coefficient agree perfectly with the single channel two-potential model.  
The small but non-vanishing coefficient of $1/x_{max}$ in the BC24$^*$ case provides an estimate for the differences between the  single channel analytical approach, that includes the adiabatic plus non-adiabatic contributions, and the multi-channel model with a full numerical calculation. The BC24$^*$ boundary condition yields the smallest coefficient of the $1/x_{max}$ term in $\mathcal M(x_{max})$, see~\autoref{tab:fits}, which is 
approximately equal to $0.198/20 \sim 0.01$~ru for $\mathcal I=20$~ru and $x_{max}=20$~ru.
This value gives an estimate of the error introduced in the 
field-dressed scattering length when the dependence of $\mathcal M(x_{max})$ on $x_{max}$ is neglected and when the field-dressed scattering length is set equal to $\mathcal M(x_{max})$ instead of $\mathcal M_0$.  
The boundary conditions BC24$^*$ thus represent the best choice to systematically calculate the field-dressed scattering length.

The coefficient of $1/x_{max}^2$ in the numerical expansion of $\mathcal M(x_{max})$, labeled $\eta({\mathcal I}, x_{00})$, 
see column 6 of~\autoref{tab:fits}, depends on $\mathcal I$ and on $x_{00}$, and have a dependence on $x_{00}$  similar to the one of ${\mathcal M}_0({\mathcal I}, x_{00})=\widetilde {a}({\mathcal I}, x_{00})$. One has 
\begin{equation}
\eta({\mathcal I},x_{00})=m_2 \widetilde {a}({\mathcal I}, x_{00}) + m_1
\label{eq:eta}
\end{equation}
with coefficients depending both on $\mathcal I$ and on the boundary conditions.
The two-potential model predicts also a linear transformation, 
with coefficients depending on the reference functions and on the coefficients $c_4$, $c_{4f}$ and $c_5$,
see~\autoref{tab:M(x)}. 
We find that the slope $m_2$  is proportional to $\mathcal I^2$ and the additional 
constant $m_1$ to $ \mathcal I^3$. This could be expected from the analytical results of \autoref{tab:M(x)} and from the 
adiabatic potential~\eqref{eq:anal-ad} and the non-adiabatic coupling term~\eqref{eq:anal-n-ad}, whose  
$1/x^4$ and $1/x^5$ terms vary as $\mathcal I^2 $ and $ \mathcal I^3$, respectively.

%
%*******************************************************************************
%-------------------------------------------------------------------------------
%%%%%%%%%%%%%%%%%%%%%%%%%%%%% fig4 %%%%%%%%%%%%%%%%%%%%%%%%%%%%%%%%%%%%%%%
%
%%%%%%%%%%%%%%%%%%%%%%%%%%%%%%%%%%%%%%%%%%%%%%%%%%%%%%%%%%%%%%%%%%%%%%%%%%%%%%%%
%
The field-dressed scattering length as function of $x_{00}$
is presented in~\autoref{fig:scatt-dim} for  $\mathcal I=10$~ru, comparing $n=2, 3$ and 4 channels. 
The considered range of $x_{00}$ corresponds to one quasi-period of
the field-free scattering length, \ie the latter, which  is also
plotted (in gray), varies once over the whole domain [$-\infty$, $+\infty$]. When increasing $n$ by one unit, an additional divergence appears, related to the new channel with a larger $\ell$. 
While the poles with  $\tilde\ell=0$ and $\tilde\ell=2$ are difficult to distinguish, because the coupling between these two channels is very large at this intensity, it is easy to identify the $\tilde\ell \ge 4$ value of the field-dressed channel associated with a particular divergence, and to 
see how  the position of these divergences changes with increasing $n$.
The location of the poles with $\tilde\ell=0$ and $\tilde\ell=2$ 
are approximately stabilized as soon as the $\ell=4$ and $\ell=6$ channels are introduced in the model.
Indeed, the field-dressed scattering length reaches an almost stable value in the 3-channel model, except for divergences related to shape resonances with $\tilde\ell \ge 4$.
The width of the poles, as  $x_{00}$ is varied, is large for $\tilde\ell$=0 and 2, and becomes narrower as $\tilde\ell$ increases.
By fixing  $x_{00}$ and varying $\mathcal I$, the width of the poles
as a function of intensity decreases rapidly as $\tilde\ell$ increases, see~\autoref{fig:scatt-zi} and~\autoref{fig:SLcontrol}. 
In this range of intensity, a control of the scattering length could only use the $\tilde\ell$=0 and 2 divergences.  

%---------------------------------------------------------------------------------------
\bibliography{shaperes}
%\input{AM-SL.bbl}

%

%---------------------------------------------------------------------------------------
\end{document}